\documentclass[iop,apj]{emulateapj}
\usepackage{amsmath,natbib}
\usepackage{graphics,graphicx}

\usepackage{apjfonts}

\input{buote.defs}

\newcommand\src{\hbox{{RXJ~1159+5531}}}
\newcommand\planck{{\sl Planck}}

\begin{document} 

\title{The Entire Virial Radius of the Fossil Cluster RXJ1159+5531:
  II. Dark Matter and Baryon Fraction}
\author{David A.\ Buote\altaffilmark{1}, Yuanyuan Su\altaffilmark{1,2},
Fabio Gastaldello\altaffilmark{3}, \& Fabrizio Brighenti\altaffilmark{4}}
\altaffiltext{1}{Department of Physics and Astronomy, University of
California at Irvine, 4129 Frederick Reines Hall, Irvine, CA 92697-4575}
\altaffiltext{2}{Harvard-Smithsonian Center for Astrophysics, 60 Garden Street, Cambridge, MA 02138}
\altaffiltext{3}{INAF-IASF-Milano, Via E. Bassini 15, I-20133 Milano, Italy}
\altaffiltext{4}{Dipartimento di Fisica e Astronomia, Universita di Bologna, via Ranzani 1, 40126 Bologa, Italy}

\slugcomment{Accepted for Publication in The Astrophysical Journal}

\begin{abstract}

  In this second paper on the entire virial region of the relaxed
  fossil cluster \src\ we present a hydrostatic analysis of the
  azimuthally averaged hot intracluster medium (ICM) using the results
  of Paper~1~\citep{su15a}.  For a model consisting of ICM, stellar
  mass from the central galaxy (BCG), and an NFW dark matter (DM)
  halo, we obtain a good description of the projected radial profiles
  of ICM emissivity and temperature that yield precise constraints on
  the total mass profile. The BCG stellar mass component is clearly
  detected with a $K$-band stellar mass-to-light ratio, $M_{\star}/L_K
  = 0.61\pm 0.11\, M_{\odot}/L_{\odot}$, consistent with stellar
  population synthesis models for a Milky-Way IMF. We obtain a halo
  concentration, $c_{200}=8.4\pm 1.0$, and virial mass,
  $M_{200}=(7.9\pm 0.6)\times 10^{13}\, M_{\odot}$. For its mass, the
  inferred concentration is larger than most relaxed halos produced in
  cosmological simulations with \planck\ parameters, consistent with
  \src\ forming earlier than the general halo population. The baryon
  fraction at $r_{200}$, $f_{\rm b,200}=0.134\pm 0.007$, is slightly
  below the \planck\ value (0.155) for the universe. However, when we
  take into account the additional stellar baryons associated with
  non-central galaxies and the uncertain intracluster light (ICL),
  $f_{\rm b,200}$ increases by $\approx 0.015$, consistent with the
  cosmic value and therefore no significant baryon loss from the
  system. The total mass profile is nearly a power law over a large
  radial range ($\sim 0.2$-10~$R_e$), where the corresponding density
  slope $\alpha$ obeys the $\alpha-R_e$ scaling relation for massive
  early-type galaxies. Performing our analysis in the context of MOND
  still requires a large DM fraction ($85.0\% \pm 2.5\%$ at
  $r=100$~kpc) similar to that obtained using the standard Newtonian
  approach. The detection of a plausible stellar BCG mass component
  distinct from the NFW DM halo in the total gravitational potential
  suggests that $\sim 10^{14}\, M_{\odot}$ represents the mass scale
  above which dissipation is unimportant in the formation of the
  central regions of galaxy clusters.

\end{abstract}


\section{Introduction}
\label{intro}

It is well appreciated that galaxy clusters are powerful tools for
cosmological studies, especially through their halo mass function and
global baryon fractions~\citep[e.g.,][]{alle11a,krav12a}. In order to obtain
mass measurements of ever larger numbers of cluster masses at higher
redshifts, studies must resort to global scaling relations, often
involving proxies for the mass. Global scaling relations of ICM
properties are particularly useful to probe cooling and feedback in
cluster evolution. Measurements of global scaling relations must be
interpreted within the context of a general paradigm (e.g.,
$\Lambda$CDM) that makes definite assumptions about the full radial
structure of a halo. It is essential that those assumptions be
verified for as many systems as possible through detailed radial
mapping of halo properties.

Fortunately, there are several powerful probes of cluster mass
distributions (e.g., galaxy kinematics, ICM temperature and density
profiles, gravitational lensing, SZ effect) which, ideally, can be
combined to achieve the most accurate picture of cluster
structure~\citep[e.g.,][]{reip13a}. In practice, different techniques
are better suited for particular clusters because of multiple factors,
such as distance and mass.

 Well-known advantages of studying the ICM include (1) that it traces
 the three-dimensional cluster potential well; (2) the electron mean
 free path is sufficiently short (especially when considering the
 presence of weak magnetic fields) to guarantee the fluid
 approximation holds (i.e., with an isotropic pressure tensor); and
 (3) the hydrostatic equilibrium approximation should apply within the
 virialized region, allowing the gravitating mass to be derived
 directly from the temperature and density profiles of the
 ICM~\cite[e.g.,][]{sara86a,etto13a}. 

 Since clusters are still forming in the present epoch, deviations
 from the hydrostatic approximation are expected. Cosmological
 simulations expect typically 10\%-30\% of the total ICM pressure is
 non-thermal, primarily arising from random turbulent
 motions~\citep[e.g.,][]{rasi04a,naga07a,ecke16a}. 
 Even with the very unfortunate demise of {\sl Astro-H},
   eventually microcalorimeter detectors will provide for the first
 time precise direct measurements of ICM kinematics for many bright
 clusters, greatly reducing (or eliminating) this greatest source of
 systematic error in ICM studies~\citep[e.g.,][]{kita14a}. Even so,
 for the most reliable hydrostatic analysis it is desirable that the
 correction for non-thermal pressure be as small as possible; i.e.,
 for the most relaxed systems.

 It turns out to be difficult to find clusters with undisturbed ICM
 within their entire virial region. The clusters that tend to be the
 most dynamically relaxed over most of their virial region are the
 cool core clusters which, unfortunately, are also those that most
 often display ICM disturbances in their central regions believed to
 arise from intermittent feedback from an AGN in the central
 galaxy~\citep[e.g.,][]{byko15a}. Hence, CC clusters with the least
 evidence for central ICM disturbance are probably the best clusters
 for hydrostatic studies.

The cool core fossil cluster \src\ is especially well-suited for
hydrostatic studies of its ICM. It is both sufficiently bright and
distant $(z=0.081)$ allowing for its entire virial region to be mapped
through a combination of \chandra\ and \suzaku\ observations with
feasible total exposure time. The high-quality \chandra\ image reveals a highly
regular ICM with no evidence of large, asymmetrical disturbances
anywhere, including the central regions~\citep{hump12a}.  We have
\suzaku\ observations covering the entire region within $r_{200}$ on
the sky and presented results for the ICM properties in each of four
directions (including using the central \chandra\ observation)
in~\citet[][hereafter Paper~1]{su15a}. We found the ICM properties
(e.g, temperature, density) to display only very modest azimuthal
variations ($<10\%$), providing evidence for a highly relaxed ICM.

In this second paper on the entire virial radius of \src\ we focus on
measurements of the BCG stellar mass-to-light ratio, dark matter, gas,
and baryon fraction.  Whereas Paper~1 focused on the comparison of ICM
properties obtained for each of the four directions observed by
\suzaku, here we analyze results for all the directions together to
obtain the best-fitting azimuthally averaged ICM and mass
properties. For calculating distances we assumed a flat $\Lambda$CDM
cosmology with $\Omega_{m,0}=0.3$ and $H_0 =
70$~km~s$^{-1}$~Mpc$^{-1}$.  At the redshift of \src\ ($z=0.081$) this
translates to an angular-diameter distance of $315$~Mpc and
$1\arcsec=1.5$~kpc. Unless stated otherwise, all statistical
uncertainties quoted in this paper are $1\sigma$.

The paper is organized as follows. We review very briefly the
observations, data preparation, and ICM measurements in \S
\ref{obs}. We discuss our implementation of the hydrostatic method in
\S \ref{method} and define the particular models and parameters in \S
\ref{mod}.  We present the results in \S \ref{results} and describe
the construction of the  systematic error budget in \S \ref{sys}.
Finally, in \S \ref{disc} we discuss several implications of our
results and present our summary and conclusions in \S \ref{conc}. 

\section{Observations}
\label{obs}

The observations and data preparation are reported in Paper~1, and we
refer the interested reader to that paper for details. Briefly, after
obtaining cleaned events files for each observation, we extracted
spectra in several concentric circular annuli for each data set and
constructed appropriate response files for each annulus; i.e.,
redistribution matrix files (RMFs) and auxiliary response files
(ARFs, including ``mixing'' ARFs to account for the large,
energy-dependent Point Spread Function of the \suzaku\ data). We
constructed model \suzaku\ spectra representing the ``non-cosmic''
X-ray background (NXB) and subtracted these from the observations.
All other background components for both the \chandra\ and \suzaku\
data were accounted for with simple parameterized models fitted
directly to the observations using \xspec\ v12.7.2~\citep{xspec}; i.e., the
annular spectrum of each data set was fitted individually with a
complex spectral model consisting of components for the ICM and
background. For each annulus on the sky a single thermal plasma
component ({\sc vapec}) was fitted to represent all the ICM emission in that
annulus. No deprojection was performed during the spectral fitting
because it amplifies noise and renders the analysis of the
background-dominated cluster outskirts even more
challenging. Furthermore, standard deprojection algorithms (e.g.,
onion-peeling and {\sc projct} in \xspec) do not generally account for
ICM emission outside the bounding annulus, which also can lead to
sizable systematic effects~\citep[e.g.,][]{nuls95a,mcla99a,buot00c}.

Consequently, the principal data products resulting from our analysis
are the radial profiles of projected (1) emission-weighted temperature
and (2) emissivity-weighted $\rho_{\rm gas}^2$; e.g., equations B10
and B13 of \citet{gast07b}. In addition, we also use the profile of
projected, emission-weighted iron abundance ($Z_{\rm Fe}$) expressed
in solar units~\citep{aspl} to further constrain the emissivity in our
hydrostatic models.

\section{Entropy-Based Method}
\label{method}

We prefer to construct hydrostatic equilibrium models using an
approach that begins by specifying a parameterized model for the ICM
entropy~\citep{hump08a}. The benefits of this ``entropy-based''
approach, as well as a review of other methods, are presented in
\citet{buot12a}. Compared to the temperature and density, the
entropy profile is more slowly varying and has a well-motivated
asymptotic form, $\sim r^{1.1}$ for all
clusters~\citep[e.g.,][]{tozz01a,voit05a}. In addition, by requiring
the entropy to be a monotonically increasing function of radius, the
additional contraint of convective stability (not typically applied in
cluster mass studies) is easily enforced. We assume spherical symmetry
which, if in fact the cluster is a triaxial ellipsoid, introduces only
modest biases into the inferred parameters (see \S \ref{sys.sphere}).

For studies of cluster ICM the thermodynamic entropy is usually
replaced by the entropy proxy, $S \equiv k_{\rm B}Tn_e^{-2/3},$
expressed in units of keV~cm$^{2}$.  It is useful to define the
quantity, $S_{\rho} \equiv (k_{\rm B}/\mu m_{\rm a})T\rho_{\rm
  gas}^{-2/3},$ where $\mu$ is the mean atomic mass of the ICM and
$m_a$ is the atomic mass unit, that replaces $n_e$ in the entropy
proxy with $\rho_{\rm gas}$~\citep[e.g., using eqn.\ B4 of][]{gast07b}
so that,
\begin{equation}
\frac{S_{\rho}}{S} = \frac{1}{\mu m_{\rm
    a}}\left(\frac{2+\mu}{5\mu}\frac{1}{m_{\rm a}}\right)^{2/3}.
\end{equation}
The equation of hydrostatic equilibrium may now be written,
\begin{equation}
\frac{d\xi}{dr} = -\frac{2}{5}\frac{GM(<r)}{r^2}S_{\rho}^{-3/5}
\label{eqn.he}
\end{equation}
where $\xi \equiv P^{2/5}$, $P$ is the total thermal pressure, and
$M(<r)$ is the total gravitating mass enclosed within radius
$r$. Given $S_{\rho}$ (after specifying $S$) and $M(<r)$, the
hydrostatic equation can be integrated directly to obtain $\xi$ and
therefore the profiles of gas density, $\rho_{\rm gas}=
(P/S_{\rho})^{3/5} = S_{\rho}^{-3/5}\xi^{3/2}$, and temperature,
$k_{\rm B}T/\mu m_{\rm a} = S_{\rho}^{3/5}P^{2/5} =
S_{\rho}^{3/5}\xi$. By comparing the density and temperature profiles
to the observations, we constrain the parameters of the input $S$ and
$M(<r)$ models.

Since $M(<r)$ in eqn.\ (\ref{eqn.he}) contains $M_{\rm gas}$, direct
integration only yields a self-consistent solution for $\rho_{\rm
  gas}$ provided $M_{\rm gas}\ll M(<r).$ In Paper~1, and all our
previous studies of the entropy-based method, we insured
self-consistency in the case where $M_{\rm gas}$ cannot be neglected
by differentiating the equation with respect to r and making use of
the equation of mass continuity; e.g., see eqn.\ (4) of Paper~1. Here
we instead employ an iterative solution of eqn.\ (\ref{eqn.he}) by
treating $M_{\rm gas}$ as a small perturbation. We solve eqn.\
(\ref{eqn.he}) initially by setting $M_{\rm gas}=0$. From this
solution we use the new $\rho_{\rm gas}$ to compute the profile of
$M_{\rm gas}$. We then insert it into the hydrostatic equation and
obtain an improved solution. The process is repeated until the value
of $M_{\rm gas}$ near the virial radius changes by less than a desired
amount.

A boundary condition on $\xi$ must be specified to obtain a unique
solution of eqn.\ (\ref{eqn.he}). We choose to specify the
``reference pressure,'' $P_{\rm ref}$, at a radius of 10~kpc. Hence, the
free parameters in our hydrostatic model are $P_{\rm ref}$ and those
associated with $S$ and $M(<r)$, which we detail below in \S
\ref{mod}.

To compare to the observations, the three-dimensional density and
temperature profiles of the ICM obtained from eqn.\ (\ref{eqn.he}) are
used to construct the volume emissivity,
$\epsilon_{\nu}\propto\rho_{\rm gas}^2 \Lambda_{\nu}(T,Z)$, where
$\Lambda_{\nu}(T,Z)$ is the ICM plasma emissivity. Then
$\epsilon_{\nu}$ and the emission-weighted temperature are projected on
to the sky and averaged over the various circular annuli corresponding
to the X-ray data using equations B10 and B13 of \citet{gast07b}.
Since $\Lambda_{\nu}(T,Z)$ also depends on the metal abundances, we
need to specify the abundance profiles in our models. As in our
previous studies, we obtain best results by simply using the measured
abundance profiles in projection and assigning them to be the true
three-dimensional profiles for the models. (The iron abundance
profiles are presented in Fig.~3 of \citealt{hump12a} and Fig.~13 of
Paper~1.) In \S \ref{mischm} we discuss instead using a parameterized
model for the iron abundance that is emission-weighted and projected
onto the sky and fitted to the observations.

Since the X-ray emission in each annulus on the sky generally contains
the sum of a range of temperatures and metallicities owing to radial
gradients in the ICM properties, the values in particular of the
temperature and metallicity obtained by fitting single-component ICM
models to the projected spectra can be substantially biased with respect to the
emission-weighted values~\citep[e.g.,][]{buot00a,mazz04a}. To
partially mitigate such biases in the data-model comparison, we employ
``response weighting'' of our projected models~\citep[see eqn.\ B15
of][]{gast07b}. 

\section{Models and Parameters}
\label{mod}

\subsection{Entropy}

For the ICM entropy profile we employ a power-law with two breaks
plus a constant,
\begin{equation}
S(r) = s_0 + s_1f(r), \label{eqn.entropy}
\end{equation}
where $s_0$ represents a constant entropy floor, and $s_1=S(r_{\rm
  ref}) - s_0$ for some reference radius $r_{\rm ref}$ (taken to be
10~kpc). The dimensionless function $f(r)$ is,
\begin{equation*}
f(r) = \left \{
\begin{array}{l}
\left(\frac{r}{r_{\rm ref}}\right)^{\alpha_1}  \hfill r \le r_{\rm b,1}\\
f_1\left(\frac{r}{r_{\rm ref}}\right)^{\alpha_2}  \qquad\qquad \hfill r_{\rm b,1}< r \le r_{\rm b,2}\\
f_2\left(\frac{r}{r_{\rm ref}}\right)^{\alpha_3}   \hfill r > r_{\rm b,2}
\end{array}  \right .\
\end{equation*}
where $r_{\rm b,1}$ and $r_{\rm b,2}$ are the two break radii, and the
coefficients $f_1$ and $f_2$ are given by,
\begin{equation*}
f_n  =  f_{n-1} \left(\frac{r_{{\rm b},n}}{r_{\rm ref}}\right)^{\alpha_{n}-\alpha_{n+1}},
\end{equation*}
with $f_0\equiv 1.$ To enforce convective stability (\S \ref{method}) we
require $\alpha_1,\alpha_2,\alpha_3\ge 0.$ Hence, this model has seven free parameters:
$s_0$, $s_1$, $r_{\rm b,1}$, $r_{\rm b,2}$, $\alpha_1$, $\alpha_2$,
and $\alpha_3$.

\subsection{Stellar Mass}

We represent the stellar mass of the cluster using the $K$-band light
profile of the BCG (2MASX J11595215+5532053) from the Two Micron
All-Sky Survey (2MASS) as listed in the Extended Source
Catalog~\citep{jarr00a}; i.e., an $n=4$ Sersic model (i.e., de Vaucouleurs)
with $r_e=9.83$~kpc and $L_K=1.03\times 10^{12}\,
L_{\odot}$. Additional (poorly constrained) stellar mass contributions
from non-central galaxies and intracluster light are treated as a
systematic error in \S \ref{sys.stars}.

\subsection{Dark Matter}

We consider the following models for the distribution of dark matter. 

\begin{itemize}
\item{\bf NFW} As it is the current standard both for modeling
  observations and simulated clusters, we use the NFW profile~\citep{nfw} for our
  fiducial dark matter model. It has two free parameters, a
  concentration $c_{\Delta}$, and mass $M_{\Delta}$, evaluated
  with respect to an overdensity $\Delta$ times the critical density
  of the universe. 
\item{\bf Einasto} The Einasto profile~\citep{eina65} is now recognized as a
  more accurate representation of the profiles of dark matter
  halos. We implement the mass profile following~\citet{merr06a} but
  using the approximation for $d_n$ given by~\citet{reta12a}. As with
  the NFW profile, we express the parameters of the Einasto model in
  terms of a virial concentration, $c_{\Delta}$, and mass,
  $M_{\Delta}$. We fix $n=5\, (\alpha=0.2)$ appropriate for cluster
  halos~\citep[e.g.,][]{dutt14a}. 
\item{\bf CORELOG} To provide a strong contrast to the NFW and Einasto
  models, we investigate a model having a constant density core
  and with a density that approaches $r^{-2}$ at large radius. For
  consistency, we also express the free parameters of this model in
  terms of a virial concentration, $c_{\Delta}$, and mass,
  $M_{\Delta}$; e.g., see \S 2.1.2 of \citet{buot12c} for more details.
\end{itemize}

All $c_{\Delta}$, $M_{\Delta}$, and $r_{\Delta}$ values are evaluated
at the redshift of \src~(i.e., $z=0.081$). Below we quote
concentration values for the total mass profile (i.e., stars+gas+DM)
as, $c_{\Delta}^{\rm tot}\equiv r_{\Delta}^{\rm tot}/r_s^{\rm DM}$,
where $r_s^{\rm DM}$ is the DM scale radius and $r_{\Delta}^{\rm tot}$
is the virial radius of the total mass profile. We determine
$r_{\Delta}^{\rm tot}$ iteratively starting with the DM virial radius,
adding in the baryon components, recomputing the virial radius, and
stopping when the change in virial radius is less than a desired
tolerance.

\section{Results}
\label{results}

\begin{figure*}[t]
\begin{center}
\includegraphics[scale=0.49,angle=-90]{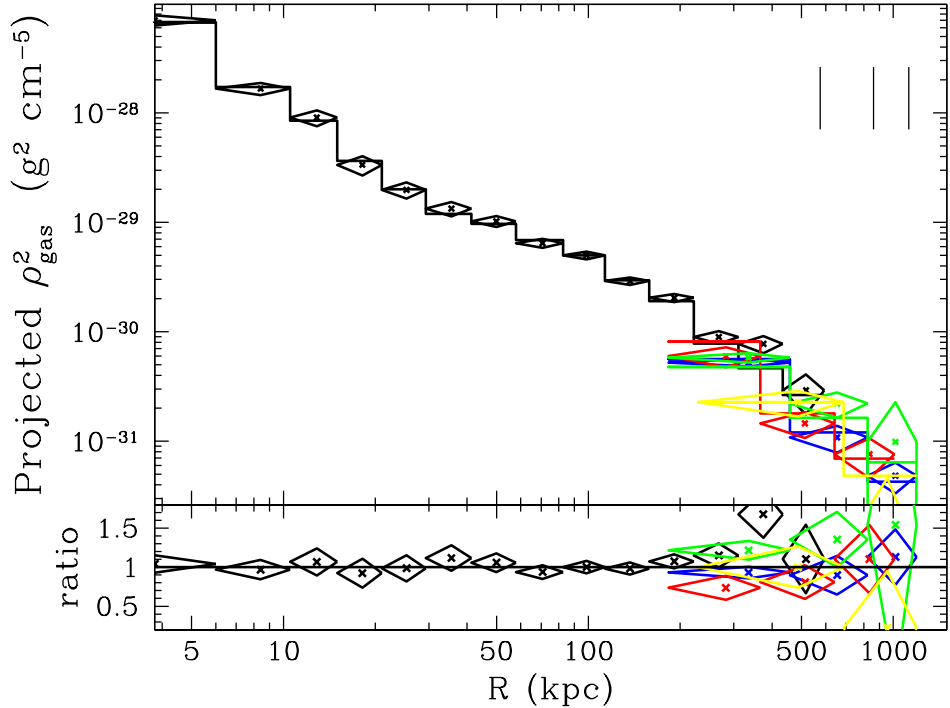}
\end{center}
\caption{\footnotesize The quantity plotted (with
    $1\sigma$ errors) is proportional to the ICM surface brightness
    divided by $\Lambda_{\nu}(T,Z)$ (see equation B10 of
    \citealt{gast07b}). The best-fitting fiducial hydrostatic model
  is also over-plotted so that its color reflects its value for each
  data set. The data sets are labelled as follows: \chandra\ (black);
  \suzaku\ pointings: N (blue), S (red), W (green), E (yellow). The
  vertical lines in the top right corner indicate the best-fitting
  virial radii for the fiducial model; i.e., from left to right:
  $r_{500}$, $r_{200}$, and $r_{108}$. The bottom panel plots the
  data/model ratio.}
\label{fig.density}
\end{figure*}

\begin{figure*}[t]
\begin{center}
\includegraphics[scale=0.49,angle=-90]{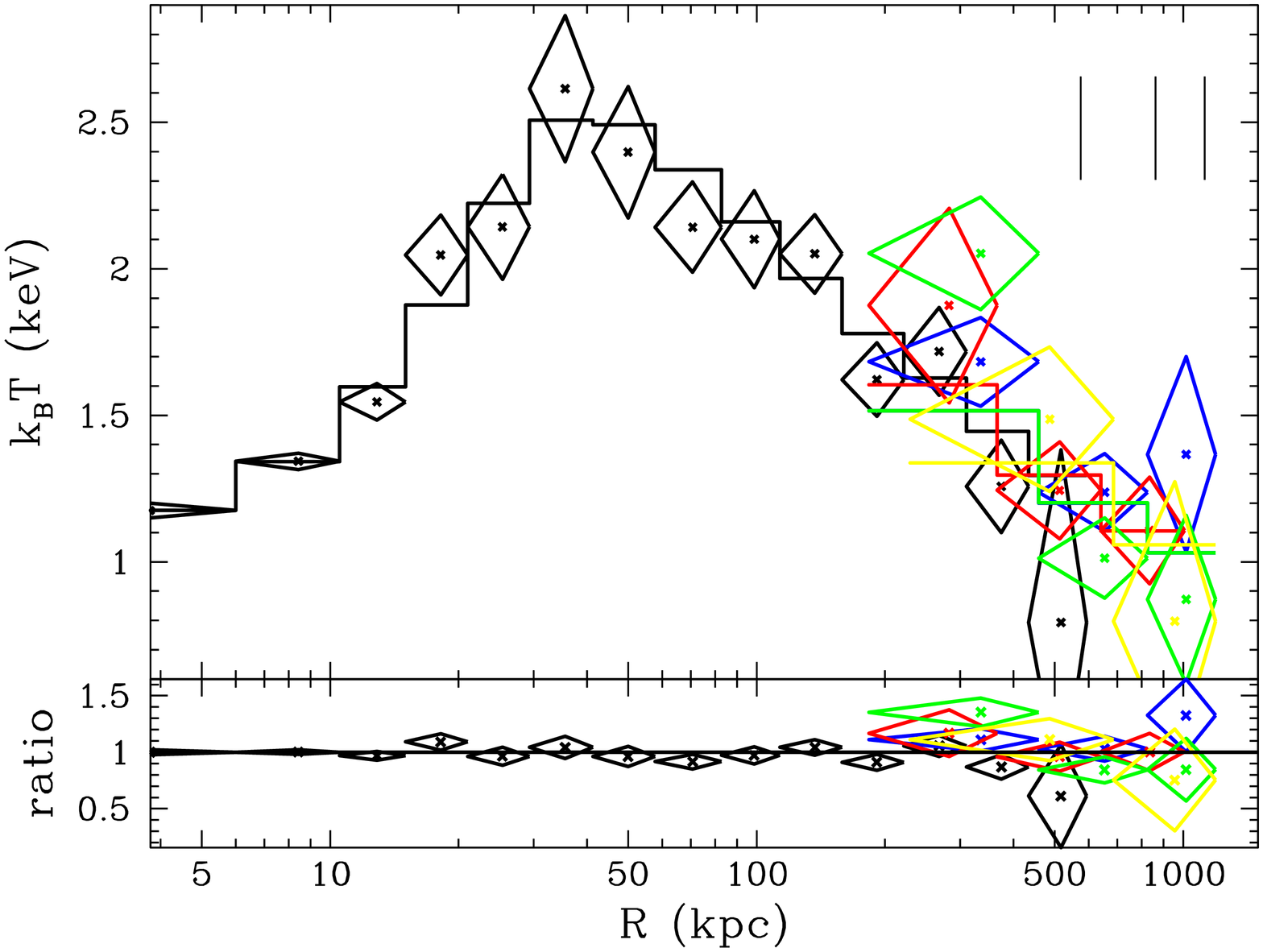}
\end{center}
\caption{\footnotesize Projected emission-weighted hot gas temperature
  ($k_BT$) along with the best-fitting fiducial hydrostatic model in
  each circular annulus versus radius $R$ on the sky. The labeling
  scheme is the same as Fig.\ \ref{fig.density} }
\label{fig.temperature}
\end{figure*}

\subsection{Overview}

We fitted the model to the data using the ``nested sampling'' Bayesian
Monte Carlo procedure implemented in the MultiNest code
v2.18~\citep{multinest}, and we adopted flat priors on the logarithms
of all the free parameters. We used a $\chi^2$ likelihood function,
where the $\chi^2$ consists of the temperature and projected
$\rho_{\rm gas}^2$ data points, the model values, and the statistical
weights. The weights are the variances of the data points obtained
from the spectral fitting in Paper~1. We quote two ``best'' values for
each parameter: (1) ``Best Fit'', which is the expectation value of
the parameter in the derived posterior probability distribution, and
(2) ``Max Like'', which is the parameter value that gives the maximum
likelihood (found during the nested sampling). Finally, unless stated
otherwise, all errors quoted are $1\sigma$ representing the standard
deviation of the parameter computed in the posterior probability
distribution. All models shown in the figures have been evaluated
using the ``Max Like'' values.

In Figures~\ref{fig.density} and \ref{fig.temperature} we display the
projected $\rho_{\rm gas}^2$
($\propto\Sigma_{\nu}/\Lambda_{\nu}(T,Z)$, where $\Sigma_{\nu}$ is the
surface brightness) and temperature data along with the best-fitting
``fiducial'' model (and residuals). The fiducial model consists of an
entropy profile with two breaks, the $n=4$ sersic model for the
stellar mass of the BCG, and the NFW model for the dark
matter. Inspection of the figures reveals that, overall, the fit is
good. Most of the fit residuals are within $\approx 1\sigma$ of the
model values, and the most deviant points lie within $\approx 2\sigma$
of the model values. Since for our Bayesian analysis we cannot easily
formally assess the goodness-of-fit as in a frequentist approach, we
have also fitted models to the data using a standard frequentist
$\chi^2$ analysis. For our fiducial model we obtain a minimum
$\chi^2=39.9$ for 39 degrees of freedom, which is formally acceptable
from the frequentist perspective. (For reference, if the BCG component
is omitted, $\chi^2=59.6$; i.e., the data strongly require it.)

Below we shall often refer to the ``Best Fit'' virial radii
of the fiducial model: $r_{2500}=271$~kpc, $r_{500}=575$~kpc,
$r_{200}=862$~kpc, and $r_{\rm vir}= r_{108}=1.12$~Mpc.

\subsection{Entropy}

\begin{figure*}
\parbox{0.49\textwidth}{
\centerline{\includegraphics[scale=0.42,angle=0]{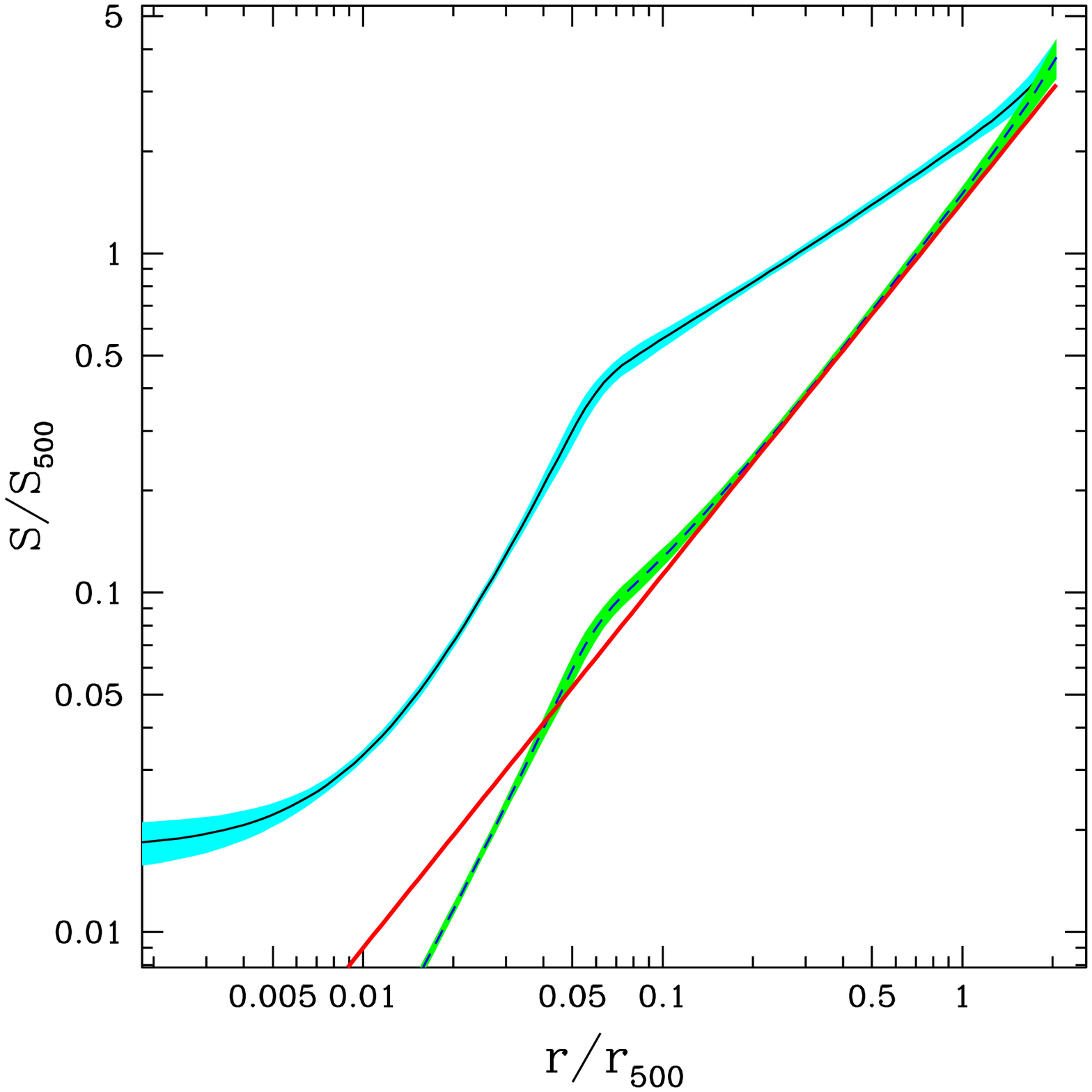}}}
\parbox{0.49\textwidth}{
\centerline{\includegraphics[scale=0.42,angle=0]{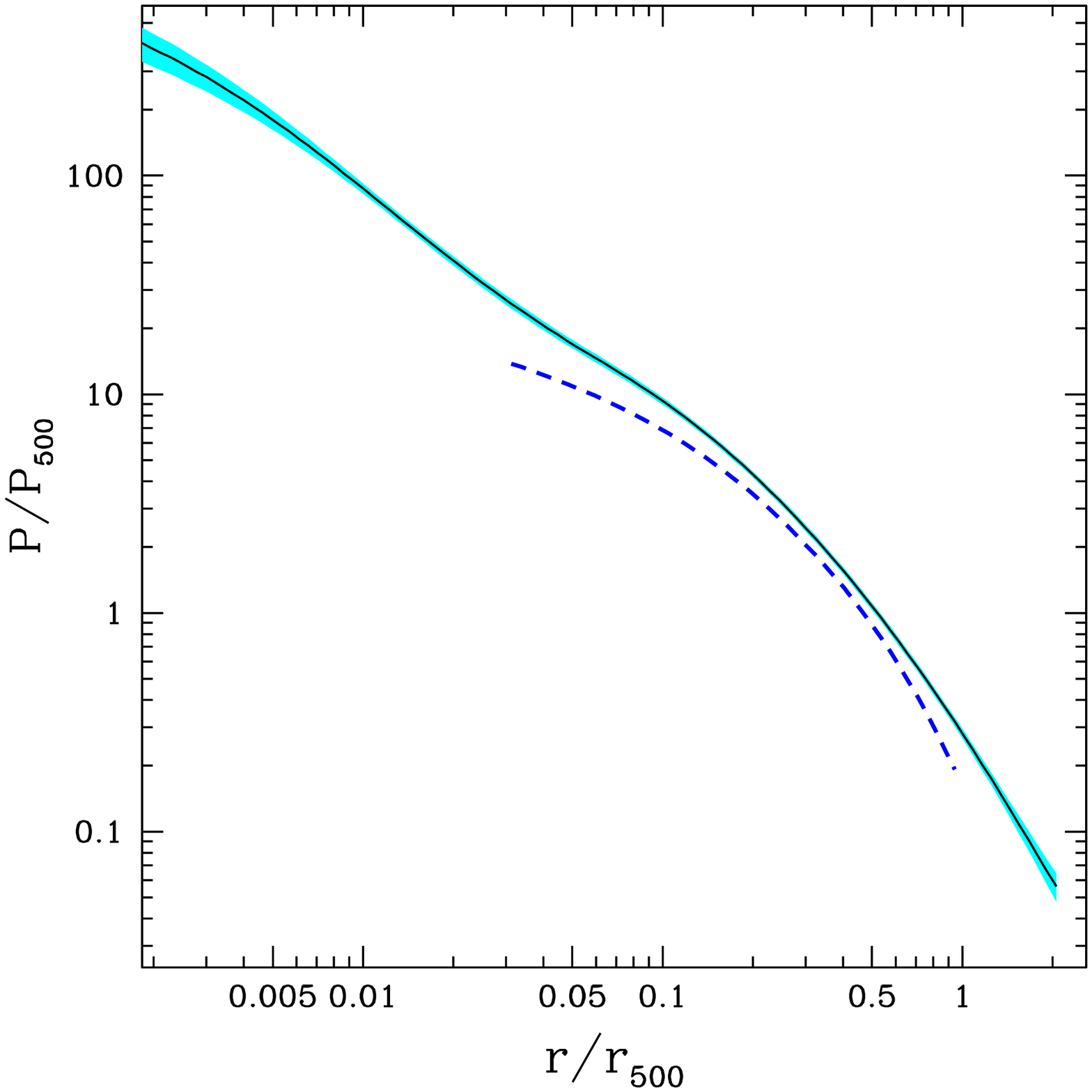}}}
\caption{\label{fig.entropy} ({\sl Left Panel}) Radial entropy profile
  (black) and $1\sigma$ error region (cyan) for the fiducial
  hydrostatic model expressed in terms of the characteristic value
  $S_{500}$. The red line indicates the baseline $r^{1.1}$ profile
  produced by gravity-only cosmological
  simulations~\citep{voit05a}. The black dashed line (and green
  $1\sigma$ region) represents rescaling the entropy profile by a
  function of the gas fraction according to \citet{prat10a}. ({\sl
    Right Panel}) Pressure profile (black) compared to the universal
  profile inferred for groups and clusters by \citet{arna10a} which is
  expected to apply for the radii shown.}
\end{figure*}

\begin{table*}[t] \footnotesize
\begin{center}
\caption{Pressure and Entropy}
\label{tab.entropy}
\begin{tabular}{lc|ccccccc}  \hline\hline\\[-7pt]
 & $P_{\rm ref}$ & $s_0$ & $s_1$ & $r_{\rm b,1}$ & $r_{\rm b,2}$ & $\alpha_1$ & $\alpha_2$ & $\alpha_3$\\ 
& ($10^{-2}$ keV cm$^{-3}$) & (keV cm$^2$) & (keV cm$^2$) & (kpc) & (kpc) \\ 
\hline \\[-7pt]
Best Fit & $5.30 \pm 0.25$ & $4.48 \pm 0.78$ & $10.9 \pm  1.3$ & $35.6 \pm  5.1$ & $ 708 \pm  277$ & $1.81 \pm 0.18$ & $0.57 \pm 0.06$ & $1.01 \pm 0.44$ \\
(Max Like) & $(5.49)$ & $(4.51)$ & $(10.7)$ & $(42.4)$ & $( 180)$ & $(1.80)$ & $(0.36)$ & $(0.74)$ \\
\hline \\
\end{tabular}
\tablecomments{Constraints on the free parameters associated with the
 pressure and entropy for the fiducial hydrostatic model. $P_{\rm
    ref}$ is the normalization of the total thermal ICM pressure profile at the
  reference radius 10~kpc. The other parameters refer to the double
  broken power-law entropy model (eqn.\ \ref{eqn.entropy}). In the top
  row we list the marginalized ``best fitting'' values and $1\sigma$
  errors; i.e., the mean values and standard deviations of the
  parameters of the Bayesian posterior. In the second row
  we give the ``maximum likelihood'' parameter values that maximize
  the likelihood function.}
\end{center}
\end{table*}

The results for the entropy profile using the fiducial hydrostatic
model are displayed in Fig.\ \ref{fig.entropy}, and the parameter
constraints are listed in Table \ref{tab.entropy}. In the figure we
have plotted the ``scaled'' entropy in units of the characteristic entropy,
$S_{500}=260.6$~keV~cm$^{2}$~\citep[see eqn.\ 3 of][]{prat10a}. 
The entropy profile has a small, but significant floor at the center
and then rises more steeply than the $\sim r^{1.1}$ profile out to the
first break radius ($\approx 36$~kpc). The profile is then shallower
than the baseline model out to the second break ($\approx
700$~kpc), after which the slope is uncertain, but consistent with the
baseline model out to the largest radii investigated $(\sim r_{\rm 108})$. 
As noted in Paper~1, at no radius does the scaled entropy fall below the
baseline model, consistent with a simple feedback explanation. (See
Paper~1 for more detailed discussion of how the entropy profile
compares to theoretical models.)

In Fig.\ \ref{fig.entropy} following \citet{prat10a} we also show the
result of rescaling the entropy profile by
$(f_{\rm gas}/f_{b,U})^{2/3}$, where $f_{\rm gas}$ is the gas
fraction as a function of radius and $f_{b,U}=0.155$ is the baryon
fraction of the Universe. The overall very good agreement of this
rescaled entropy profile with the baseline model suggests that the
feedback has primarily served to spatially redistribute the gas rather than
raise its temperature. 

We note that the second break in the entropy profile is not required
at high significance; i.e., the ratio of the Bayesian evidences for the
2-break and 1-break model is 1.9. 

\subsection{Pressure}

The results for the pressure profile using the fiducial hydrostatic
model are displayed in Fig.\ \ref{fig.entropy}, and the constraints
for the reference pressure are listed in Table \ref{tab.entropy}. In
the figure we have plotted the ``scaled'' pressure in units of the
characteristic pressure, $P_{500}=5.9\times
10^{-4}$~keV~cm$^{-3}$~\citep[see eqn.\ 5 of][]{arna10a}, and compared
to the ``universal'' pressure profile of \citet{arna10a}. (Note that
we quote results for the total gas pressure rather than the electron
gas pressure and have accounted for this also in the definition of
$P_{500}$.) In most of the region where the universal profile is
expected to be valid, the pressure profile of \src\ agrees within the
20\% scatter in the pressure profiles of the clusters studied by
\citet{arna10a}, reaching maximum deviations of 50-60\% at the
endpoints; i.e., \src\ has a pressure profile similar to the other
clusters.

\subsection{Mass}
\label{mass}

\begin{figure*}
\parbox{0.49\textwidth}{
\centerline{\includegraphics[scale=0.42,angle=0]{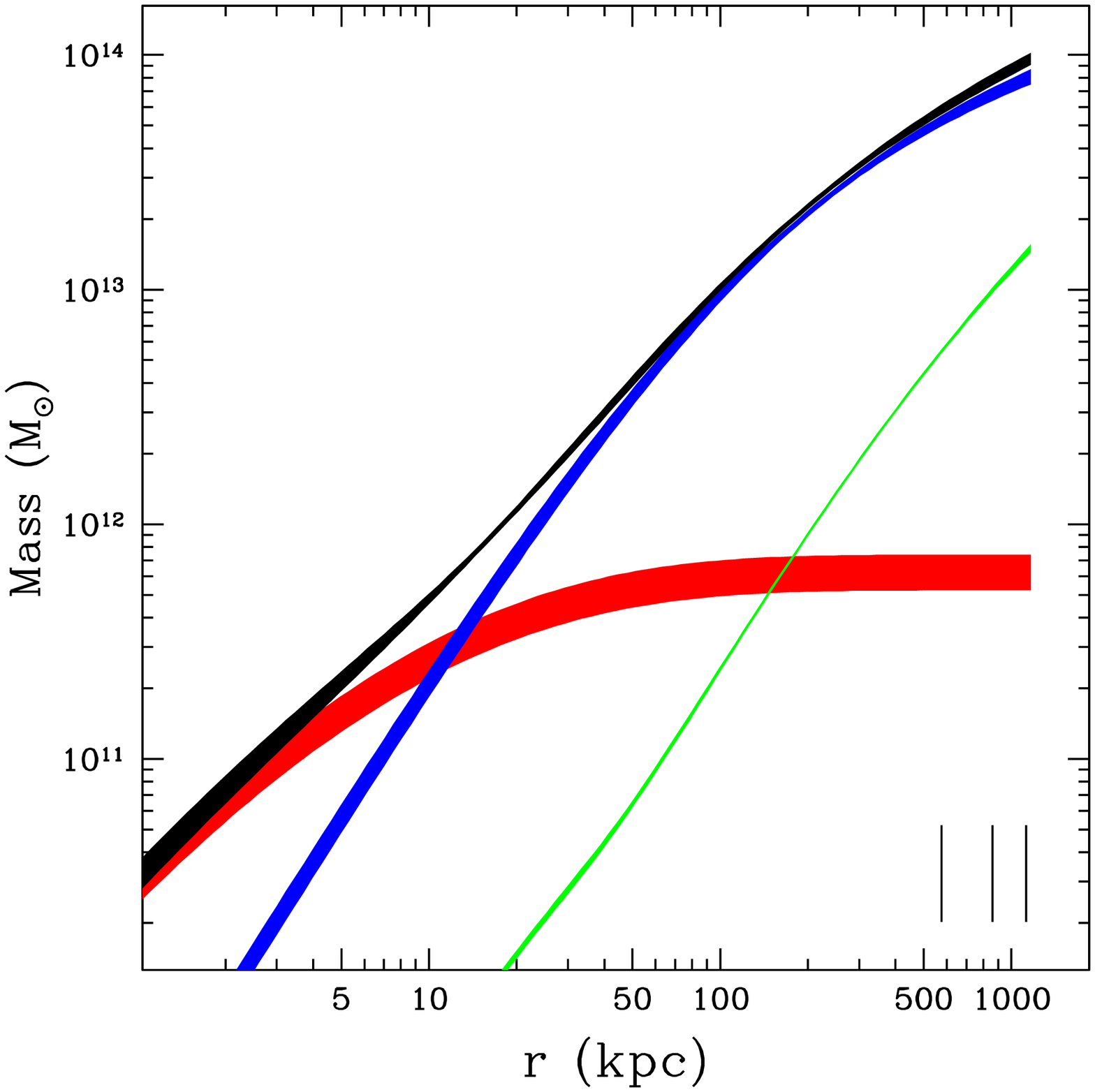}}}
\parbox{0.49\textwidth}{
\centerline{\includegraphics[scale=0.42,angle=0]{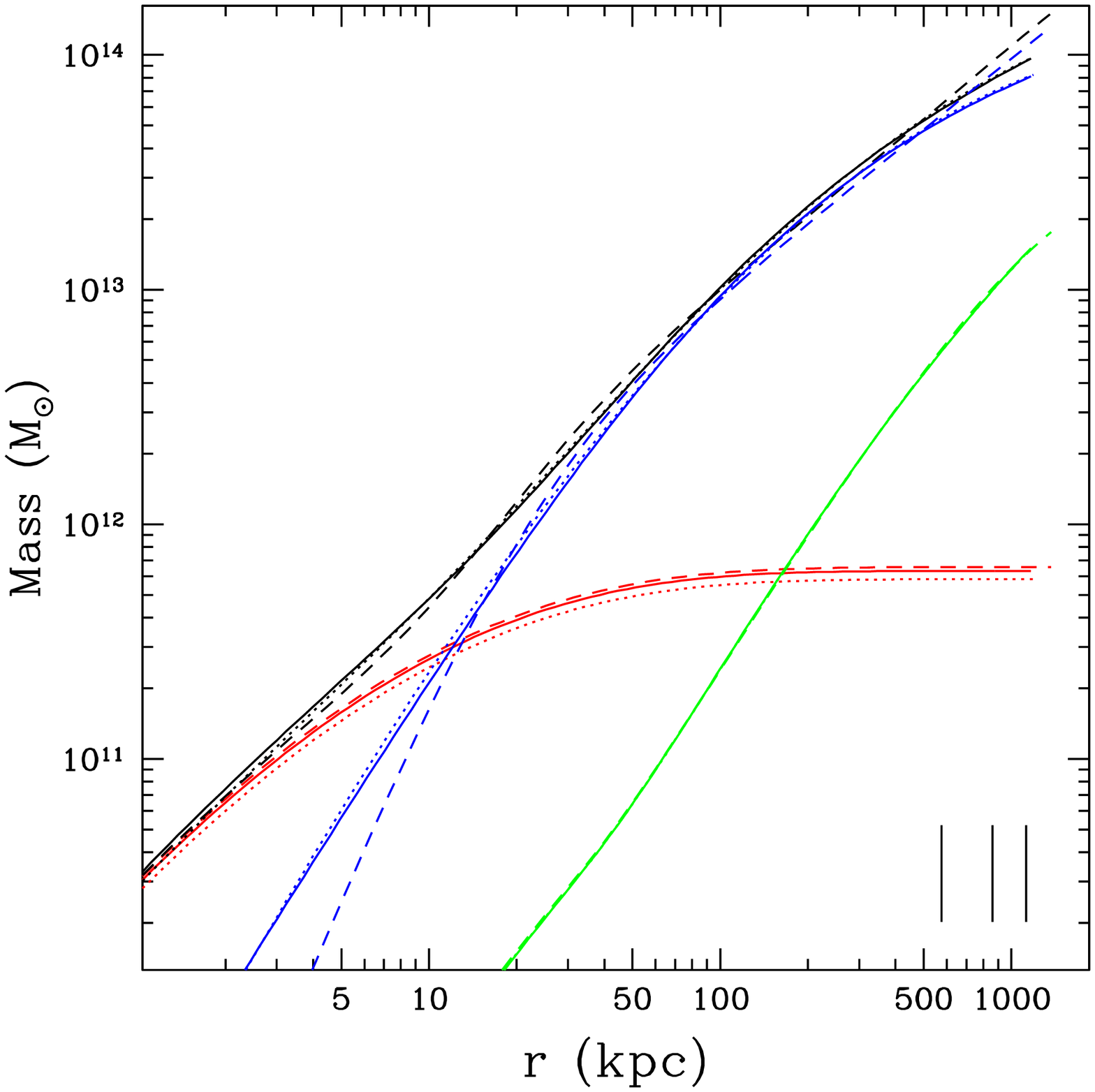}}}
\caption{\label{fig.mass} ({\sl Left Panel}) Radial profiles of the
  various mass components of the fiducial hydrostatic model: total
  mass (black), NFW DM (blue), stars (red), hot gas (green). The
  vertical lines in the bottom right corner indicate the best-fitting
  virial radii for the fiducial model; i.e., from left to right:
  $r_{500}$, $r_{200}$, and $r_{108}$. ({\sl Right Panel}) Results
  shown for hydrostatic models having different dark matter profiles:
  NFW (solid lines), Einasto (dotted lines), Corelog (dashed
  lines). Only the best-fitting models are shown. The color scheme is
  the same as for the left panel. The gas mass profiles are
  indistinguishable between the models.}
\end{figure*}

\renewcommand{\arraystretch}{1.5}

\begin{table*}[t] \footnotesize
\begin{center}
\caption{Stellar and Total Mass}
\label{tab.mass}
\begin{tabular}{lc|cc|cc|cc|cc}  \hline\hline\\[-7pt]
& $M_{\star}/L_K$ & $c_{2500}$ & $M_{2500}$  & $c_{500}$ & $M_{500}$  & $c_{200}$ & $M_{200}$  & $c_{\rm vir}$ & $M_{\rm vir}$\\ 
& ($M_{\odot} L_{\odot}^{-1}$) &  & ($10^{13}M_{\odot}$) &  & ($10^{13}M_{\odot}$) &  & ($10^{13}M_{\odot}$) &  & ($10^{13}M_{\odot}$)\\
\hline \\[-7pt]
Best Fit & $0.61 \pm 0.11$  & $ 2.6 \pm  0.4$ & $ 3.1 \pm  0.1$ & $ 5.6 \pm  0.7$ & $ 5.9 \pm  0.4$ & $ 8.4 \pm  1.0$ & $ 7.9 \pm  0.6$ & $10.9 \pm  1.3$ & $ 9.4 \pm  0.7$\\ 
(Max Like) & $(0.53)$ & $( 3.0)$ & $( 3.1)$ & $( 6.3)$ & $( 5.6)$ & $( 9.4)$ & $( 7.4)$ & $(12.2)$ & $( 8.8)$\\ 
\hline \\[-7pt]
Spherical & $\cdots$ & $\cdots$ & $\cdots$ & $_{-0.32}^{+0.13}$ & $_{-0.04}^{+0.15}$ & $\cdots$ & $\cdots$ & $\cdots$ & $\cdots$\\
Einasto & $-0.05$ & $-0.2$ & $-0.0$ & $-0.4$ & $ 0.1$ & $-0.6$ & $ 0.1$ & $-0.9$ & $ 0.1$ \\ 
CORELOG & $-0.05$ & $10.0$ & $-0.4$ & $23.1$ & $ 0.5$ & $37.5$ & $ 2.4$ & $51.7$ & $ 4.7$ \\ 
1 Break & $0.02$ & $-0.1$ & $ 0.0$ & $-0.2$ & $ 0.1$ & $-0.2$ & $ 0.2$ & $-0.3$ & $ 0.2$ \\ 
$\Lambda_{\nu}(T,Z)$ & $-0.04$ & $ 0.2$ & $-0.1$ & $ 0.5$ & $-0.4$ & $ 0.7$ & $-0.5$ & $ 0.9$ & $-0.6$ \\ 
Response & $0.02$ & $-0.2$ & $ 0.0$ & $-0.4$ & $ 0.1$ & $-0.6$ & $ 0.3$ & $-0.8$ & $ 0.3$ \\ 
Proj. Limit & $^{+0.00}_{-0.00}$ & $^{+ 0.0}_{-0.0}$ & $^{+ 0.0}_{-0.0}$ & $^{+ 0.0}_{-0.0}$ & $^{+ 0.0}_{-0.0}$ & $^{+ 0.1}_{-0.0}$ & $^{+ 0.0}_{-0.0}$ & $^{+ 0.1}_{-0.0}$ & $^{+ 0.0}_{-0.0}$ \\ 
SWCX & $0.00$ & $-0.0$ & $ 0.0$ & $-0.0$ & $ 0.0$ & $-0.1$ & $ 0.1$ & $-0.1$ & $ 0.1$ \\ 
Distance & $^{+0.03}_{-0.02}$ & $^{+ 0.1}_{-0.1}$ & $^{+ 0.0}_{-0.0}$ & $^{+ 0.1}_{-0.2}$ & $^{+ 0.1}_{-0.1}$ & $^{+ 0.2}_{-0.3}$ & $^{+ 0.1}_{-0.1}$ & $^{+ 0.2}_{-0.3}$ & $^{+ 0.2}_{-0.2}$ \\ 
$N_{\rm H}$ & $-0.02$ & $ 0.1$ & $-0.0$ & $ 0.2$ & $-0.1$ & $ 0.3$ & $-0.2$ & $ 0.4$ & $-0.3$ \\ 
Fix $Z_{\rm Fe}(r_{\rm out})$ & $^{+0.05}_{-0.03}$ & $^{+ 0.2}_{-0.3}$ & $^{+ 0.2}_{-0.1}$ & $^{+ 0.3}_{-0.5}$ & $^{+ 0.5}_{-0.2}$ & $^{+ 0.5}_{-0.8}$ & $^{+ 0.7}_{-0.3}$ & $^{+ 0.6}_{-1.1}$ & $^{+ 0.8}_{-0.4}$ \\ 
Solar Abun. & $-0.05$ & $ 0.3$ & $-0.1$ & $ 0.5$ & $-0.3$ & $ 0.7$ & $-0.5$ & $ 1.0$ & $-0.6$ \\ 
PSF & $0.03$ & $-0.1$ & $ 0.1$ & $-0.3$ & $ 0.2$ & $-0.4$ & $ 0.3$ & $-0.5$ & $ 0.4$ \\ 
FI-BI & $0.02$ & $-0.1$ & $-0.0$ & $-0.1$ & $ 0.0$ & $-0.2$ & $ 0.1$ & $-0.2$ & $ 0.1$ \\ 
NXB & $^{+0.01}_{-0.00}$ & $-0.0$ & $^{+ 0.1}_{-0.0}$ & $-0.0$ & $^{+ 0.1}_{-0.0}$ & $-0.1$ & $ 0.1$ & $-0.1$ & $^{+ 0.1}_{-0.0}$ \\ 
CXB & $-0.01$ & $ 0.0$ & $ 0.0$ & $ 0.1$ & $-0.0$ & $ 0.1$ & $-0.0$ & $ 0.1$ & $-0.1$ \\ 
CXBSLOPE & $^{+0.06}_{-0.04}$ & $^{+ 0.2}_{-0.4}$ & $^{+ 0.3}_{-0.1}$ & $^{+ 0.4}_{-0.7}$ & $^{+ 0.8}_{-0.4}$ & $^{+ 0.7}_{-1.0}$ & $^{+ 1.1}_{-0.5}$ & $^{+ 0.9}_{-1.4}$ & $^{+ 1.2}_{-0.6}$ \\ 
\hline \\
\end{tabular}
\tablecomments{Best-fitting values, maximum likelihood values, and
  $1\sigma$ errors (see notes to Table \ref{tab.entropy}) for the
  stellar mass-to-light ratio $(M_{\star}/L_K)$, concentration, and
  enclosed mass corresponding to the total mass profile (stars+gas+DM)
  of the fiducial hydrostatic model computed for radii corresponding
  to several different over-densities. (Note: ``vir'' refers to
  $\Delta=108.)$ We also provide a detailed systematic error budget as
  explained in \S \ref{sys}.  For most of the entries we list the
  values using the same precision. If the particular error has a
  smaller value, it appears as a zero; e.g., ``0.0'' or
  ``-0.0''. Briefly, the systematic test names refer to tests of, 
\\
(``Spherical'', \S \ref{sys.sphere}): the assumption of spherical symmetry\\
(``Einasto'', ``CORELOG'',  \S \ref{sys.dm}): using Einasto or CORELOG instead of NFW for the
  DM profile\\
(``1 Break'', \S \ref{sys.entropy}): only using one break radius for the
  entropy model\\
(``$\Lambda_{\nu}(T,Z)$'', \S \ref{mischm}): different
  treatments for deprojecting the plasma emissivity\\
(``Response'', \S \ref{mischm}): response weighting\\
(``Proj.\ Limit''), \S \ref{mischm}): outer radius of model cluster\\
(``SWCX'', \S \ref{bkg}): Solar Wind Charge Exchange emission\\
(``Distance'',  \S \ref{mischm}): assumed distance\\
(``$N_{\rm H}$'', \ref{miscfit}): assumed Galactic hydrogen column density\\
(``Fix $Z_{\rm Fe}(r_{\rm out})$'', \S \ref{abun})): effect of fixing the metal
abundance in the outer \suzaku\ aperture to either $0.1Z_{\odot}$ or $0.3Z_{\odot}$\\
(``Solar Abun.\", \S \ref{abun}):  using different solar abundance tables\\
 (``PSF'', \S \ref{miscfit}): the sensitivity to the PSF mixing procedure for the \suzaku\ data\\
(``FI-BI'', \S \ref{miscfit}): the flux calibration of the different \suzaku\ CCDs\\
(``NXB'', \S \ref{bkg}): the sensitivity to the adopted model of the
non-X-ray background for the \suzaku\ data\
(``CXB'', \S \ref{bkg}): fixing the normalization of the CXB power\\ law
to the average value determined from surveys\\ 
 (``CXBSLOPE'', \S \ref{bkg}): the sensitivity to the assumed slope in
 the CXB power law}
\end{center}
\end{table*}

\renewcommand{\arraystretch}{1}

In Fig.\ \ref{fig.mass} we show the profiles for the total mass and
different mass components for the fiducial hydrostatic model, while in
Table~\ref{tab.mass} we list the results for $M_{\star}/L_K$ and the NFW
parameters, concentration and mass, evaluated for several
over-densities. Of all the mass components, the stellar mass
has the weakest constraints $(\sim \pm 18\%)$, while the gas mass is
the best constrained (e.g., $\sim \pm 4\%$ at $r_{\rm 108}$.),
considering only the statistical errors. 

Most of the possible systematic errors we consider in \S \ref{sys},
and listed in Table~\ref{tab.mass}, are not significant since they
induce parameter changes of the same size or smaller than the
$1\sigma$ statistical error. The most significant changes result from
aspects of the background modeling (\S \ref{bkg}), the treatment of
the plasma emissivity (\S \ref{mischm}), and how the metal abundances
are treated (\S \ref{abun}), particularly in the outermost apertures;
i.e., the CXBSLOPE, $\Lambda_{\nu}(T,Z)$, Solar Abun, and Fix $Z_{\rm
  Fe}(r_{\rm out})$ rows in Table~\ref{tab.mass}. It is, however,
reassuring that even these changes generally lead to parameter changes
not much larger than $1\sigma$. (See \S \ref{sys} for a more detailed
discussion of the systematic error budget.)

%
%

The value we obtain for the $K$-band stellar mass-to-light ratio
($M_{\star}/L_K\approx 0.6 M_{\odot}/L_{\odot}$) of the BCG agrees
very well with our previous determinations~\citep{gast07b,hump12a} and
also is consistent with the value expected from stellar population
synthesis models. Using the published relationship between stellar
mass-to-light ratio and color from \citet{zibe09a}, we obtain
$M_{\star}/L_K\approx 0.55 M_{\odot}/L_{\odot}$ for $g-i = 1.41$,
where the $g$ and $i$ magnitudes are taken from the ``Model'' entries
in the NASA/IPAC Extragalactic Database (NED) referring to the Sloan
Digital Sky Survey Data Release
6~\footnote{http://www.sdss.org/dr6/products/catalogs/index.html}.
This good agreement should be considered only a mild consistency
check, since (1) there is significant scatter depending on which bands
are used for the color (we used $g-i$ as favored by
\citealt{zibe09a}), and (2) the relationship between color and
$M_{\star}/L_K$ depends on the assumed stellar initial mass function
(IMF). If instead we use the relationship between stellar
mass-to-light-ratio and color of \citet{bell03a}, who employ a
Salpeter-like IMF, we obtain $M_{\star}/L_K\approx 0.94
M_{\odot}/L_{\odot}$, about $3\sigma$ above our measured value. While
there is considerable scatter depending on the color used, the X-ray
analysis favors the lower $M_{\star}/L_K$ obtained
from~\citet{zibe09a} who adopt a Milky-Way
IMF~\citep{chab03a}. Several previous studies have found that massive
early-type galaxies instead favor a Salpeter
IMF~\citep[e.g.,][]{conr12a,newm13b,dutt14b}, although the more recent
study by~\citep[e.g.][]{smit15a} and most of our previous X-ray
studies~\citep[e.g.,][]{hump09c,hump12b} favor a Milky-Way IMF.

Since fossil clusters like \src\ are thought to be highly evolved,
early forming systems, it is interesting to examine whether they
possess large concentrations for their mass compared to the general
population. According to the results of \citet{dutt14a}, the mean
value of $c_{200}$ for a ``relaxed'' dark matter halo with
$M_{200}=7.9\times 10^{13}\, M_{\odot}$ in the \planck\ cosmology is
5.3, which is a little more than $3\sigma$ lower than the value we
measure for \src\ $(8.4\pm 1.0)$. In addition, with respect to the
intrinsic scatter of the theoretical relation, the $c_{200}$ value is
a $\sim 2\sigma$ outlier. Hence, with respect to the fiducial
hydrostatic model with an NFW DM halo, \src\ appears to possess a
significantly above-average concentration, consistent with forming
earlier than the average halo population.  For the Einasto DM halo,
our measured $c_{200}=7.8\pm 1.0$ is slightly more than $2\sigma$
above the mean predicted value of 5.6 and a little above the $1\sigma$
intrinsic scatter ($c_{200}=7.5$) using the results of
\citet{dutt14a}. The reduced significance for the Einasto case arises
from a combination of our smaller measured $c_{200}$ and also larger
theoretical intrinsic scatter for the Einasto profile compared to NFW.

In the right panel of Fig.\ \ref{fig.mass} we compare the results for
the different dark matter models. As is readily apparent from visual
inspection, the gas mass profiles are virtually identical for all
three cases; i.e., the inferred gas mass profile is robust to the
assumed dark matter model. Over most of the radii investigated, the
total mass profile is also quite insensitive to the assumed dark
matter model, though CORELOG leads to a more massive halo than either
NFW or Einasto. (Note in particular the ``pinching'' of the total mass
profile near 10-20~kpc, about 1-2~$R_e$, where the dark matter crosses
over the stellar mass.)  The largest deviations appear at the very
largest radii ($\gtrsim r_{108}$).  The fitted stellar mass profile is
very similar for all three DM models and yields consistent
$M_{\star}/L_K$ values in each case.

All the DM models produce fits of similar quality in terms of
the magnitude of their fractional residuals. When we perform a
frequentist $\chi^2$ fit we obtain minimum $\chi^2$ values of 39.4
for Einasto, 40.1 and for CORELOG compared to 39.9 for NFW; i.e., the
fits are statistically indistinguishable from the frequentist
perspective. Moreover, from the Bayesian analysis we can use the ratio
of evidences to compare the Einasto $(\ln Z = -53.00)$ and NFW
$(\ln Z = -53.81)$ models since they have essentially the same
free parameters and priors: We obtain an evidence ratio of 2.2 in
favor of the Einasto model, which is not very significant.  It is not
straightforward to compare the evidences of the CORELOG and NFW models
because they have different prior volumes (and the models are not
``nested''). Nevertheless, the frequentist minimum $\chi^2$ values
clearly show that, despite the high-quality X-ray data covering the
entire virial radius in projection, the data do not statistically
disfavor the CORELOG model.

For completeness, we have also examined allowing the Einasto index $n$
to be a free parameter. Since, as just noted, the data are unable to
distinguish clearly between the NFW, Einasto $(n=5)$, and CORELOG
profiles (each of which have two free parameters), and these models
already provide formally acceptable fits, it follows that adding
another free parameter does not improve the fit very much. Indeed, for
the frequentist fit we find that $\chi^2$ is reduced by only 0.08 and
gives a large $1\sigma$ error range on the index: $n =
5.8^{+4.6}_{-2.0}$ or $\alpha = 1/n = 0.17^{+0.09}_{-0.08}$. Despite
the large uncertainty, the best-fitting Einasto index matches well the
value expected for a DM halo of the mass of \src~\citep{dutt14a}.

\subsection{Gas and Baryon Fraction}
\label{baryfrac}

\begin{figure}[t]
\begin{center}
\includegraphics[scale=0.42]{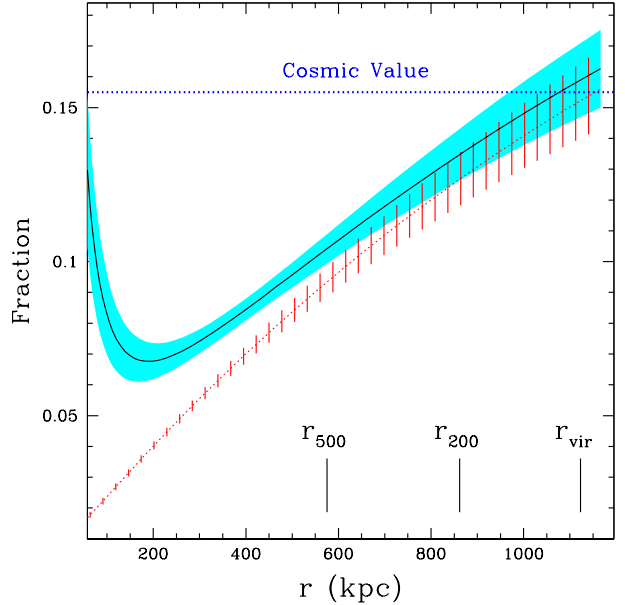}
\end{center}
\caption{\footnotesize 
Radial profiles of the baryon (solid black line, cyan $1\sigma$ error
region) and gas (dotted red line, red $1\sigma$ error region)
fractions of the fiducial hydrostatic model.}
\label{fig.baryfrac}
\end{figure}

\renewcommand{\arraystretch}{1.5}
\begin{table*}[t] \footnotesize
\begin{center}
\caption{Gas and Baryon Fraction}
\label{tab.fb}
\begin{tabular}{lcc|cc|cc|cc}   \hline\hline\\[-7pt]
& $f_{\rm gas, 2500}$ & $f_{\rm b, 2500}$ & $f_{\rm gas, 500}$ & $f_{\rm b, 500}$ & $f_{\rm gas, 200}$ & $f_{\rm b, 200}$ & $f_{\rm gas, vir}$ & $f_{\rm b, vir}$\\
\hline \\[-7pt]
 & $0.051 \pm 0.001$ & $0.072 \pm 0.004$ & $0.093 \pm 0.003$ & $0.104 \pm 0.004$ & $0.126 \pm 0.007$ & $0.134 \pm 0.007$ & $0.152 \pm 0.010$ & $0.159 \pm 0.010$\\ 
(Max Like) 
 & $(0.052)$ & $(0.070)$ & $(0.093)$ & $(0.103)$ & $(0.127)$ & $(0.134)$ & $(0.155)$ & $(0.161)$\\ 
\hline \\[-7pt]
$M_{\rm stellar}^{\rm other}$ & $\cdots$ & $^{+0.017}$ & $\cdots$ & $^{+0.016}$ &
$\cdots$ &$^{+0.016}$ & $\cdots$ & $^{+0.015}$\\
Spherical & $\cdots$ & $\cdots$ & $_{-0.0009}^{+0.0006}$ & $\cdots$ &
$\cdots$ & $\cdots$ & $\cdots$ & $\cdots$\\
Einasto & $-0.001$ & $0.000$ & $-0.002$ & $-0.001$ & $-0.002$ & $-0.001$ & $-0.002$ & $-0.001$ \\ 
CORELOG & $0.003$ & $0.002$ & $-0.003$ & $-0.001$ & $-0.018$ & $-0.015$ & $-0.037$ & $-0.035$ \\ 
1 Break & $0.000$ & $-0.000$ & $-0.001$ & $-0.001$ & $-0.001$ & $-0.001$ & $-0.000$ & $-0.000$ \\ 
$\Lambda_{\nu}(T,Z)$ & $0.000$ & $0.001$ & $0.004$ & $0.004$ & $0.010$ & $0.010$ & $0.015$ & $0.016$ \\
Response & $0.001$ & $0.001$ & $-0.001$ & $-0.001$ & $-0.003$ & $-0.003$ & $-0.004$ & $-0.004$ \\ 
Proj. Limit & $^{+0.000}_{-0.000}$ & $^{+0.000}_{-0.000}$ & $^{+0.000}_{-0.000}$ & $^{+0.000}_{-0.000}$ & $^{+0.001}_{-0.000}$ & $^{+0.001}_{-0.000}$ & $^{+0.001}_{-0.000}$ & $^{+0.001}_{-0.000}$ \\ 
SWCX & $-0.000$ & $-0.000$ & $-0.000$ & $-0.000$ & $-0.000$ & $-0.000$ & $-0.000$ & $-0.000$ \\ 
Distance & $^{+0.002}_{-0.001}$ & $^{+0.001}_{-0.000}$ & $^{+0.002}_{-0.001}$ & $^{+0.002}_{-0.001}$ & $^{+0.003}_{-0.002}$ & $^{+0.003}_{-0.001}$ & $^{+0.003}_{-0.002}$ & $^{+0.002}_{-0.002}$ \\ 
$N_{\rm H}$ & $-0.000$ & $0.000$ & $0.001$ & $0.001$ & $0.002$ & $0.002$ & $0.003$ & $0.004$ \\ 
Fix $Z_{\rm Fe}(r_{\rm out})$ & $0.000$ & $^{+0.001}_{-0.000}$ & $^{+0.003}_{-0.007}$ & $^{+0.003}_{-0.007}$ & $^{+0.006}_{-0.016}$ & $^{+0.006}_{-0.016}$ & $^{+0.009}_{-0.024}$ & $^{+0.009}_{-0.024}$ \\ 
Solar Abun. & $-0.000$ & $0.001$ & $0.004$ & $0.004$ & $0.007$ & $0.007$ & $0.008$ & $0.008$ \\ 
PSF & $0.001$ & $-0.000$ & $-0.001$ & $-0.002$ & $-0.003$ & $-0.003$ & $-0.004$ & $-0.004$ \\ 
FI-BI & $0.002$ & $0.001$ & $0.003$ & $0.003$ & $0.005$ & $0.005$ & $0.007$ & $0.007$ \\ 
NXB & $^{+0.001}_{-0.001}$ & $^{+0.001}_{-0.000}$ & $^{+0.002}_{-0.002}$ & $^{+0.001}_{-0.002}$ & $^{+0.001}_{-0.006}$ & $^{+0.001}_{-0.005}$ & $^{+0.000}_{-0.010}$ & $^{+0.000}_{-0.010}$ \\ 
CXB & $-0.001$ & $-0.000$ & $-0.001$ & $-0.001$ & $-0.002$ & $-0.002$ & $-0.004$ & $-0.004$ \\ 
CXBSLOPE & $^{+0.001}_{-0.001}$ & $^{+0.001}_{-0.001}$ & $^{+0.005}_{-0.009}$ & $^{+0.005}_{-0.008}$ & $^{+0.011}_{-0.018}$ & $^{+0.011}_{-0.018}$ & $^{+0.018}_{-0.027}$ & $^{+0.018}_{-0.027}$ \\ 
\\ 
\hline \\
\end{tabular}
\tablecomments{Best-fitting values, maximum likelihood values, $1\sigma$
  statistical errors, and estimated systematic errors (see notes to
  Table \ref{tab.entropy}) for the gas and baryon fractions of the
  fiducial hydrostatic model computed for radii corresponding to
  several different over-densities. (Note: ``vir'' refers to
  $\Delta=108.)$}
\end{center}
\end{table*}
\renewcommand{\arraystretch}{1}

The results for the baryon and gas fraction profiles using the
fiducial hydrostatic model are displayed in Fig.\ \ref{fig.baryfrac},
and the parameter constraints are listed in Table \ref{tab.fb}
including the systematic error budget. Similar to the results for the
total mass, most of the systematic errors are insignificant in the
sense that the estimated parameter changes in the gas and baryon
fraction are comparable to or less than the $1\sigma$ statistical
error. Again the most important changes occur for aspects of the
background modeling (\S \ref{bkg}) and the treatment of the plasma
emissivity (\S \ref{mischm}) and metal abundances (\S \ref{abun}),
particularly in the outermost apertures; i.e., the CXBSLOPE,
$\Lambda_{\nu}(T,Z)$, Fix $Z_{\rm Fe}(r_{\rm out})$, and Solar Abun
rows in Table~\ref{tab.fb}.  The CXBSLOPE and Fix $Z_{\rm Fe}(r)$
result in systematic errors almost $3\sigma$ in magnitude. (See \S
\ref{sys} for a more detailed discussion of the systematic error
budget.)

For most radii $f_{\rm b}<f_{\rm b,U}$, where $f_{\rm b,U}=0.155$ is
the mean baryon fraction of the universe as determined by \planck\
\citep{plan14a}. Near $r_{108}$, $f_{\rm b}$ is consistent with
$f_{\rm b,U}$, with the hot ICM consisting of 96\% of the total
baryons. These results are virtually identical to those for the
Einasto model, where $f_{\rm b}=0.157\pm 0.011$ at $r_{108}$, whereas
CORELOG has a smaller value, $f_{\rm b}=0.121\pm 0.008$

The preceding discussion considered the baryons contained in the hot
ICM and stellar baryons associated only with the BCG as determined by
the fitted $M_{\star}/L_K$. To estimate the stellar baryons from
non-central cluster members and intracluster light (ICL) we follow the
procedure we adopted in \S 4.3 of \citet{hump12a}. Since the
contribution of these non-BCG stellar baryons is not measured
directly, we list it as a systematic error in Table \ref{tab.fb} (see
\S \ref{sys.stars}). At the largest radii these baryons are expected
to increase the baryon fraction by $\sim 10\%,$ which is not much
larger than the statistical error at $r_{108}.$

\subsection{Mass and Density Slopes}
\label{slope}

\begin{figure}[t]
\begin{center}
\includegraphics[scale=0.42]{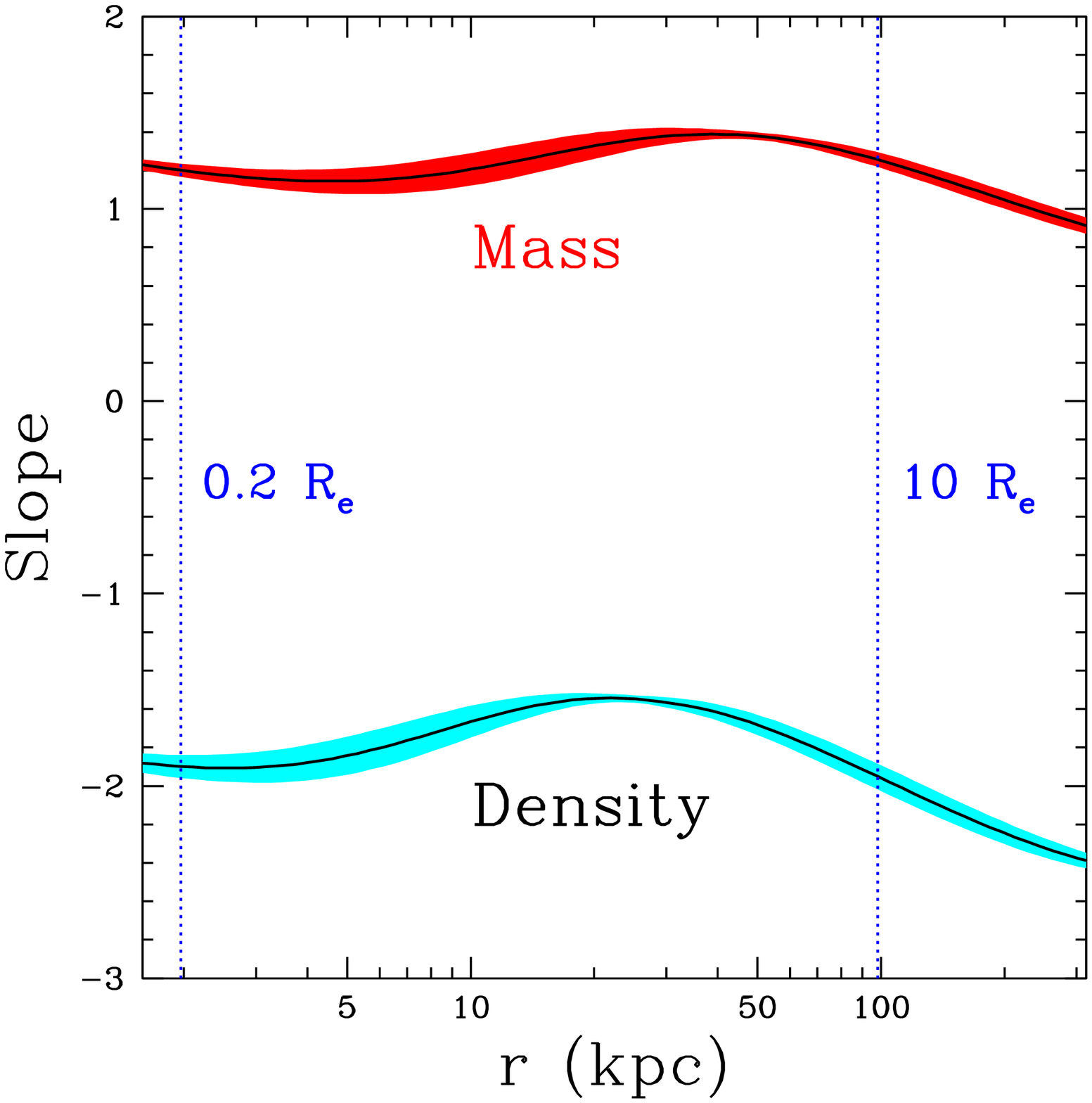}
\end{center}
\caption{\footnotesize Slopes (i.e., logarithmic derivatives) of the
  total mass and total mass density for the fiducial hydrostatic
  model. The error regions shown are $1\sigma$. The vertical lines
  indicate the regions between 0.2-10 stellar half-light radii $(R_e)$
  of the central galaxy that were studied in \citet{hump10a}.}
\label{fig.slope}
\end{figure}

\begin{table}[t] \footnotesize
\begin{center}
\caption{Mass-Weighted Total Density Slope}
\label{tab.slope}
\begin{tabular}{rrc}   \hline\hline\\[-7pt]
Radius & Radius\\
(kpc) & ($R_e$) & $\langle\alpha\rangle$\\
\hline \\[-7pt]
      4.9 	&     0.5 	  &       $1.86 	\pm         0.07$\\
      9.8 	&     1.0 	  &       $1.80 	\pm        0.08$\\
    19.7 	&     2.0 	  &       $1.67 	\pm         0.07$\\
    49.2 	&     5.0 	  &       $1.62 	\pm         0.02$\\
    98.3 	&   10.0 	  &       $1.74 	\pm         0.04$\\
\\ 
\hline \\
\end{tabular}
\tablecomments{The mass-weighted slope is computed for the fiducial
  hydrostatic model with eqn.\ \ref{eqn.slope}.}
\end{center}
\end{table}

It is now well established that the total mass profiles of massive
elliptical galaxies have density profiles very close to $\rho\sim
r^{-\alpha}$ with $\alpha\approx 2$ over a wide range in radius. This
relation extends to higher masses with smaller $\alpha$ in the central
regions of clusters~\citep[e.g.,][and references
therein]{hump10a,newm13a,cour14a,capp15a}.  In particular, in
\citet{hump10a} it was shown that the total mass profiles inferred
from hydrostatic studies of hot gas in massive elliptical galaxies,
groups, and clusters, are fairly well-approximated by a single
power-law over 0.2-10 stellar half-light radii $(R_e)$ so that
approximately, $\alpha=2.31 - 0.54\log (R_e/\rm kpc),$ a result
consistent with that obtained from combination of stellar dynamics and
strong gravitational lensing for massive elliptical
galaxies~\citep{auge10a}.

In Fig.\ \ref{fig.slope} we display the radial logarithmic derivatives
(i.e., slopes) of the total mass and total density profiles for the fiducial
hydrostatic model. The slopes are indeed slowly varying for our
models, ranging from $\approx 1.2-1.4$ for the mass to $\approx$ -1.6
to -2.0 for the density over radii 0.2-10~$R_e$, representing a radial variation of
$\pm 20\%$. 
In Table~\ref{tab.slope} we quote the mass-weighted total density
slope $\langle\alpha\rangle$ following equation (2) of \citet{dutt14b}, 
\begin{equation}
\langle\alpha\rangle = 3 - \frac{d\ln M}{d\ln r}, \hskip 0.4cm
\alpha\equiv -\frac{d\ln \rho}{d\ln r} \label{eqn.slope},
\end{equation}
where $M$ is the total mass enclosed within radius $r$. Within
$10R_e$, $\langle\alpha\rangle = 1.74\pm 0.04$, which is consistent
with the value of $\alpha=1.67^{+0.11}_{-0.10}$ we obtained previously
from a power-law fit to the mass profile~\citep{hump10a} and with
$\alpha=1.77$ obtained using the $\alpha-R_e$ scaling relation.

\subsection{MOND}

\begin{figure*}[t]
\parbox{0.49\textwidth}{
\centerline{\includegraphics[scale=0.42,angle=0]{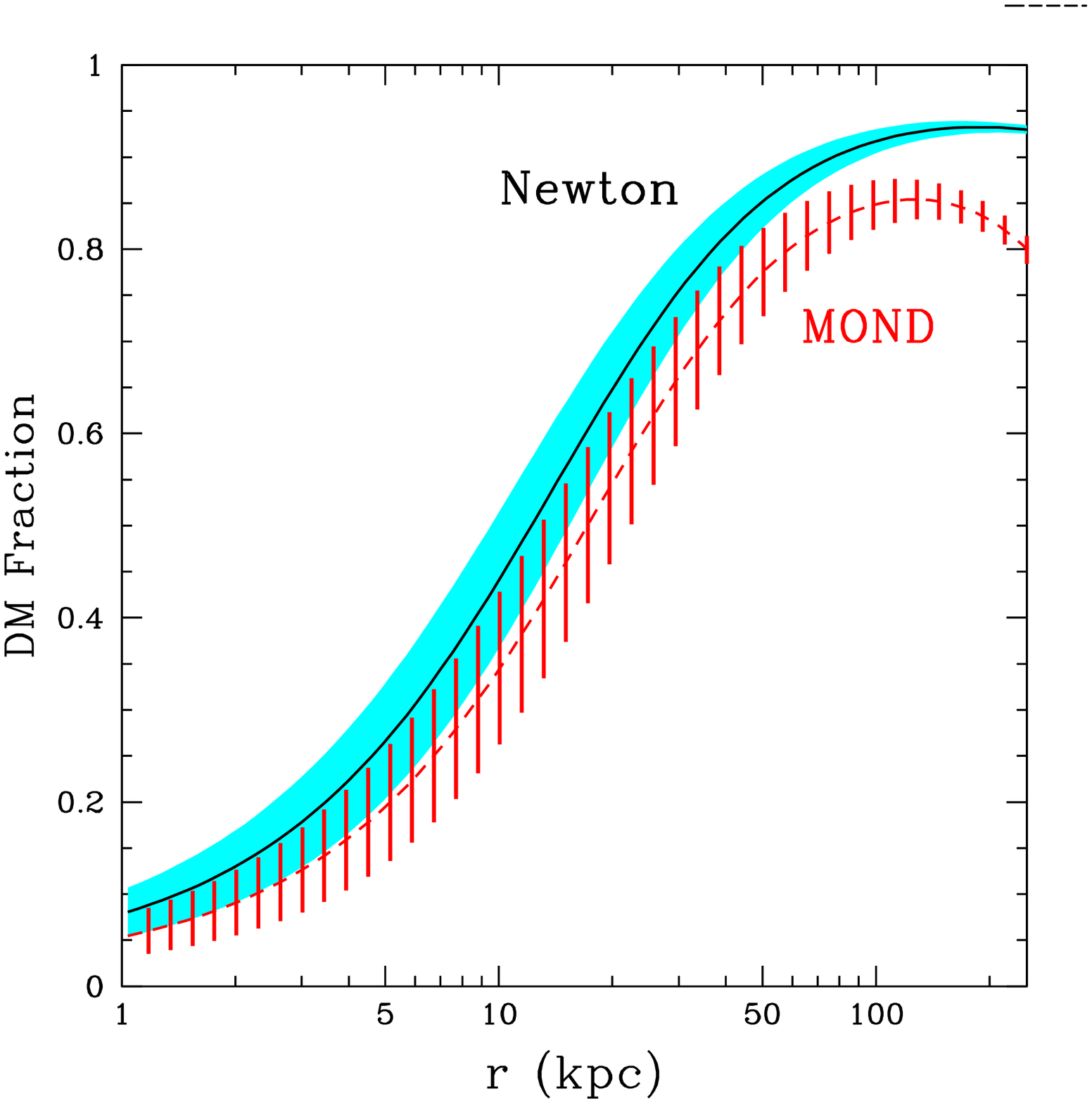}}}
\parbox{0.49\textwidth}{
\centerline{\includegraphics[scale=0.42,angle=0]{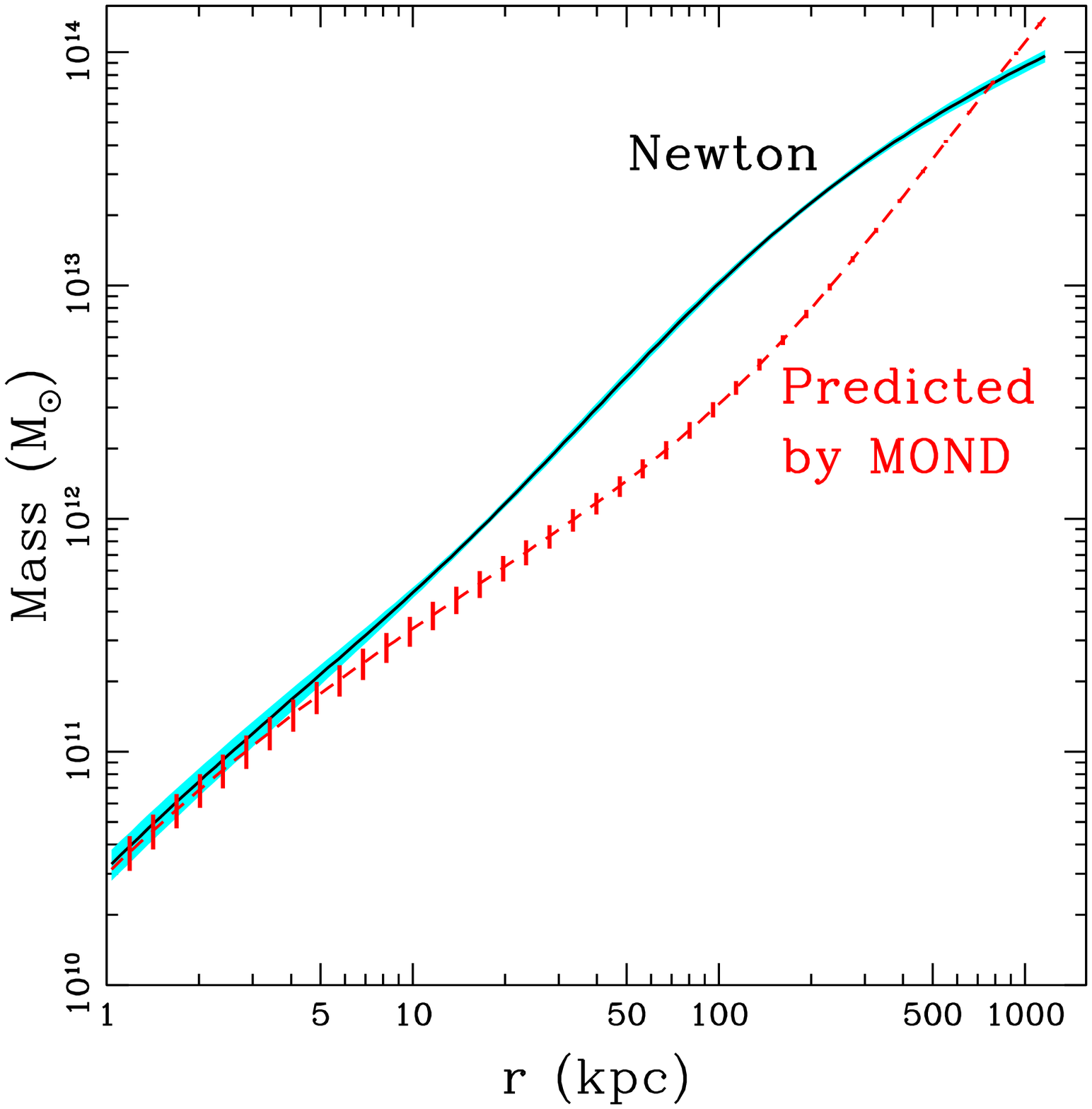}}}
\caption{\footnotesize  ({\sl Left Panel}) Radial profiles of the DM fraction for the
  fiducial hydrostatic model for the Newtonian (black and cyan) and
  MOND (red) cases. The shaded and hashed regions represent $1\sigma$
  errors. ({\sl Right Panel}) Newtonian mass profiles for (1) the total mass of the fiducial
  hydrostatic model (black and cyan, same as in Fig.\ref{fig.mass}),
  and (2) the total mass profile predicted by MOND (red) from eqn.\ \ref{eqn.mond.other}.}
\label{fig.mond}
\end{figure*}

%

While the presence of dark matter is largely accepted by the
astronomical community, it is worthwhile to examine interpretations of
the observations that instead consider a modification of the
gravitational force law. Here we consider the most widely investigated
and successful modified gravity theory, MOND~\citep{milg83a}, which
nevertheless is unable to explain observations of galaxy clusters
considering only the known baryonic
matter~\citep[e.g.,][]{sand99a,poin05a,angu08a,milg15a}. We
investigate whether MOND can obviate the need for dark matter in \src\
following the approach of \citet{angu08a}.

For an isolated
spherical system the gravitational acceleration in Newton
gravity, $g_{\rm N} = GM_{\rm N}(<r)/r^2$, and MOND, $g_{\rm M} =
GM_{\rm M}(<r)/r^2$, have similar forms, where $M_{\rm N}(<r)$ and
$M_{\rm M}(<r)$ are the respective enclosed masses within radius
$r$. For some interpolating function, $\mu(g_{\rm N}/a_0)$, the
accelerations are related by, $g_{\rm M} = \mu(g_{\rm N}/a_0) g_{\rm
  N}$, so that  $M_{\rm M} = \mu(g_{\rm N}/a_0) M_{\rm
  N}$, and $a_0\approx 1.2\times 10^{-8}$~cm~s$^{-2}$ is the MOND
acceleration constant. For $\mu(x) = x/(1+x)$,
where $x=g_{\rm N}/a_0$, we have,
\begin{equation}
M_{\rm M}(<r) = \frac{M_{\rm N}(<r)}{1 + a_0/g_{\rm N}(r)}. \label{eqn.mond}
\end{equation}
In this way we easily compute the MONDian mass using the Newtonian
mass we have derived previously; i.e., there is no different fitting
required.  Equation~(\ref{eqn.mond}), being derived from the simple
interpolating function $\mu$, has the undesirable property that
$M_{\rm M}(<r)$ reaches a maximum value at some radius and then
decreases~\citep[for more discussion of this point
see][]{angu08a}. Hence, for the moment we focus on results roughly
within the radius where the MONDian mass reaches a maximum value.

In Fig.\ \ref{fig.mond} for the fiducial hydrostatic model we compare
the cumulative dark matter fractions $(M_{\rm DM}/M_{\rm total})$
inferred from Newtonian gravity and MOND. Like in the Newtonian case,
MOND requires a dominant fraction of dark matter with increasing
radius to match the X-ray data. At $r=100$~kpc the DM fraction is
$85.0 \pm 2.5\%$; i.e., the mass discrepancy is a factor of 6.7. Even
when considering the contribution of baryons from (rather uncertain)
non-central baryons (\S \ref{sys.stars}), the DM fraction is reduced
by only a small amount to $\approx 82\%$.  

It is also interesting to view the performance of MOND from a
different perspective. Solving the MOND equation for $g_{\rm N}$ with
the same interpolating function yields, 
\begin{equation}
g_{\rm N} = \frac{g_{\rm M}}{2}\left(1 + \sqrt{1 + 4\frac{a_0}{g_{\rm
      M}}} \, \, \, \, \right). \label{eqn.mond.other}
\end{equation}
That is, given $g_{\rm M}$ evaluated using the baryonic mass profiles
(i.e., stellar and gas) derived from the Newtonian analysis, this
expression gives $g_{\rm N}$, and thus the Newtonian mass profile
(with DM), that MOND would predict. In Fig.\ \ref{fig.mond} we compare
this ``MOND predicted Newtonian mass profile'' with the actual
Newtonian total mass profile. The MOND profile under-predicts the
Newtonian mass over most radii, with the largest deficit again
occurring near $r=100$~kpc such that the predicted mass is $\approx 0.30M_{\rm
  N}$. The predicted profile crosses over the Newtonian profile shortly
before $r_{200}$ and then exceeds the Newtonian profile afterwards.



In sum, consistent with previous results for X-ray groups and clusters,
MOND requires a large fraction of dark matter similar to Newtonian
gravity to explain the X-ray data.

\subsection{Comparison to Previous Work}
\label{prev}

It is interesting to compare our results to those obtained by
\citet{hump12a} who used only the \chandra\ data and the North
\suzaku\ observation. \citet{hump12a} obtained best-fitting values,
$M_{\star}/L_K=0.54$ (in solar units), $c_{\rm vir}=11.2$, $M_{\rm vir}=9.3\times
10^{13}\, M_{\odot}$, and $f_{\rm b, 500} = 0.124$ (includes non-BCG
stellar baryons). All of these results agree within $1\sigma$ of the
values obtained in our study (Tables \ref{tab.mass} and
\ref{tab.fb}). 

The excellent agreement between the two studies is notable for several
reasons. First, the addition of the \suzaku\ observations covering out
to $r_{200}$ in the S, E, and W directions does not modify the
results significantly, consistent with the results of Paper~I indicating only small
azimuthal variation in the ICM properties at large radius. Second,
improved background modeling incorporating point sources resolved by
offset \chandra\ observations -- as well as improved calibration and
data processing in the \chandra\ and \suzaku\ pipelines -- do not
change the results significantly. Finally, the consistent results
between the studies provides a useful consistency check on the
different implementations of the entropy-based hydrostatic modeling
(e.g., treatment of self-gravity of gas mass, see \S \ref{method})
between the two studies using entirely different modeling software.


\section{Error Budget}
\label{sys}

We have considered a variety of possible sources of systematic error
and list a detailed error budget in Tables \ref{tab.mass} and
\ref{tab.fb}. Below we provide details on the construction of the error
budgets. 

\subsection{Spherical Symmetry}
\label{sys.sphere}

\citet{buot12c} showed that assuming a cluster is
spherical when in fact it is ellipsoidal does not typically introduce
large errors into the quantities inferred from hydrostatic modeling of
the ICM. For a large range of intrinsic flattenings, they computed
orientation-averaged biases (mean values and scatter) of several
derived quantities, including halo concentration, total mass, and gas
fraction. 

We use the ``NFW-EMD'' results from Table~1 of \citet{buot12c} to
provide an estimate of the error arising from the assumption of
spherical symmetry within $r_{500}$; i.e., the error from assuming the
cluster is spherical when in fact it is a flattened ellipsoid viewed
at a random orientation to the line of sight. From a study of
cosmological dark matter halos~\citet{schn12a} find that for a halo
having mass similar to \src\ the typical intrinsic short-to-long axis
ratio is $\approx 0.5.$ We adopted this value for the error estimates
on the concentration, mass, and gas fraction within $r_{500}$ in
Tables \ref{tab.mass} and \ref{tab.fb}  (``Spherical'') . In all cases the effect is
insignificant.


\subsection{Stellar Mass}
\label{sys.stars}

While the BCG dominates the stellar mass in the central region of the
cluster, smaller non-central galaxies and diffuse intracluster light
(ICL) contribute significant stellar mass at larger radius. Due to
the greater uncertainty of the amounts and distributions of these
non-central baryons, we treat their contribution as a systematic
effect to the baryon fraction. To account for these additional stellar
baryons, we follow the procedure described in \S 4.3 of
\citet{hump12a}. For the non-central galaxies we use the result of
\citet{vikh99a} that these galaxies comprise $\sim 25\%$ of the
$V$-band stellar light. We assume this result also applies in the $K$
band with the same $M_{\star}/L_K$ as the BCG. Since we do not have a
precise observational constraint on the ICL, we use the result from
the theoretical study by \citet{purc07a} that the ICL contains up to
$\sim 2$ times the stellar mass of the BCG and adopt this value to
give a conservative reflection of the systematic error. We assume  both
components of non-central baryons are spatially
distributed as the dark matter in our models.

The contribution of the non-central baryons to the baryon fraction are
listed in Table \ref{tab.fb} ($M_{\rm stellar}^{\rm other}$). These
stellar baryons increase $f_{\rm b, vir}$ by $\sim 10\%$ to $0.174$
fully consistent with the value reported in~\citet{hump12a} containing
the contributions from both the BCG and non-central stellar
baryons. This modified $f_{\rm b, vir}$ exceeds the cosmic value
$f_{\rm b,U}=0.155$ by $2\sigma$ considering only the statistical
error on our fiducial model, although the disagreement should be
considered less significant given the uncertainties in the non-central
stellar baryons. 

\subsection{Entropy Model}
\label{sys.entropy}

We examined the effect of restricting the entropy broken power-law
model to only a single break (ONEBREAK). The effect is everywhere
insignificant. 

\subsection{Dark Matter Model}
\label{sys.dm}

The effects of using a DM profile different from NFW are indicated in
the rows ``Einasto'' and ``Corelog'' in Tables \ref{tab.mass} and
\ref{tab.fb}. We have discussed the magnitudes of these systematic
differences in \S \ref{mass} and \S\ref{baryfrac}.

\subsection{Metal Abundances}
\label{abun}

We considered how choices made in the measurement of the metal
abundances from the spectral fitting affected the results. (We refer
the reader to Paper~1 for details on the spectral analysis.) First, we
examined the impact of using different solar reference
abundances. Whereas our default analysis used the solar abundance
table of \citet{aspl}, the effect of instead using the tables of
\citet{angr} or \cite{lodd} are listed in the ``Solar Abun'' row in
Tables \ref{tab.mass} and \ref{tab.fb}. The differences do not exceed
the $1\sigma$ statistical error. 

Second, since the spectra in the outermost apertures (i.e., at the
virial radius) are the most background dominated and subject to
systematic errors, we also examined the impact of fixing the metal
abundance there at $0.1Z_{\odot}$ and $0.3Z_{\odot}$, bracketing the
best-fitting value of $\sim 0.2Z_{\odot}$ (referred to as
``$\Delta$abun'' in Table~3 of Paper~1). As can be seen in the row ``Fix
$Z_{\rm Fe}(r_{\rm out})$'' in Tables \ref{tab.mass} and \ref{tab.fb},
this effect leads to one of the two largest systematic errors as
mentioned above in \S \ref{mass} and \S \ref{baryfrac}. 

\subsection{Background}
\label{bkg}

We considered the impacts of several choices made in the treatment of
the background in the spectral fitting (see Paper~1). The effect of
including a model for the solar wind charge exchange emission in the
spectral analysis is listed in row ``SWCX'' in Tables \ref{tab.mass}
and \ref{tab.fb}. We find the effect to be insignificant in all
cases. 

To assess the sensitivity of the results to the particle
background in the \suzaku\ observations, we artificially increased and
decreased the estimated non-X--ray background component by $\pm 5\%$
and list the results in row ``NXB'' in Tables \ref{tab.mass}
and \ref{tab.fb}. In all instances the differences are insignificant
for the stellar and total mass parameters. While this is also true at
most radii for the gas and baryon fractions, at  $r_{\rm vir}$ the
differences are comparable to the $1\sigma$ error. 

We also explored the sensitivity of our results to the extragalactic
Cosmic X-ray Background (CXB) power-law component (see \S 5.1 of
Paper~1). In Paper~1 by default we assumed a power-law component in
our spectral fits with a slope fixed at
$\Gamma=1.41$~\citep[e.g.,][]{delu04a} and normalization free to vary.
If we fix the normalization to the value expected for the cosmic
average (see \S 2.3 of Paper~1), we obtain the results listed in row
``CXB'' in Tables \ref{tab.mass} and \ref{tab.fb}. It is reassuring
that the differences are all negligible. If instead we keep the
normalization free to vary but change the slopes used to
$\Gamma=1.3,1.5$ we obtain the results listed in row ``CXBSLOPE''
(corresponding to ``CXB-$\Gamma$'' in Table~3 of Paper~1). As already
noted in \S \ref{mass} and \S \ref{baryfrac}, this effect leads to one
of the two largest systematic errors. The differences are comparable
to the $1\sigma$ errors within $r_{2500}$ and increase to 2-3$\sigma$
at $r_{\rm vir}$, where the data are most dominated by the background.

\subsection{Miscellaneous Spectral Fitting}
\label{miscfit}

We performed other tests associated with the spectral fitting which we
summarize here. In all cases they did not produce significant
parameter differences in Tables \ref{tab.mass} and \ref{tab.fb}. (1)
The effect of varying the adopted value of Galactic $N_{\rm H}$
\citep{dick90} by $\pm 20\%$ is shown in row ``$N_{\rm H}$'' in Tables
\ref{tab.mass} and \ref{tab.fb} (see \S 5.6 of Paper~1). (2) We varied
the spectral mixing between extraction annuli by $\pm 5\%$ from our
default case to assess the impact of small changes on how we account
for the large, energy-dependent \suzaku\ PSF (see \S 5.3 of Paper~1).
The results are given in row ``PSF'' in Tables \ref{tab.mass} and
\ref{tab.fb}. (3) By default we allowed the normalization of the ICM
model for annuli on the \suzaku\ XIS front-illuminated (FI) and
back-illuminated (BI) chips to be varied separately to allow for any
calibration differences. To assess the impact of this choice, we also
performed the spectral fitting requiring the same normalizations for
the FI and BI chips (\S 5.7 of Paper~1) and the results are given in
the row ``FI-BI'' in Tables \ref{tab.mass} and \ref{tab.fb}.

\subsection{Miscellaneous Hydrostatic Modeling}
\label{mischm}

Here we describe a few remaining tests we performed associated with
choices made in our hydrostatic modeling procedure. First, we varied
the cluster redshift by $\pm 5\%$ in our hydrostatic models (it was
similarly varied in the spectral fitting in Paper~1), and the results
are listed in row ``Distance'' of Tables \ref{tab.mass} and
\ref{tab.fb}. The differences are everywhere insignificant. Second, we
explored the impact of changing the default bounding radius of the
cluster model. Whereas the default employed is 2.5~Mpc, in row
``Proj. Limit'' of Tables \ref{tab.mass} and \ref{tab.fb} we show the
differences resulting from instead using either 2.0~Mpc or 3.0~Mpc for
the bounding radius. In all cases the effect is negligible. Next we
examined the sensitivity of the results to the ``response weighting''
(\S \ref{method}) by instead performing no such weighting. As indicated
in row ``Response'' of  Tables \ref{tab.mass} and \ref{tab.fb} the
differences are insignificant. 

Finally, we examined how our choices regarding the modeling of the
plasma emissivity affect the results.  As discussed in \S
\ref{method}, the hydrostatic models by default take the measured
metal abundance profile in projection and assigns it to be the true
three-dimensional profile which is then used (along with the
temperature) to compute the plasma emissivity
$\Lambda_{\nu}(T,Z)$. Our default procedure to do this assignment uses
the \chandra\ data and only the N \suzaku\ observation. We
investigated using instead each of the other three \suzaku\
pointings. For a more rigorous test, we also fitted a projected,
emission-weighted parametric model to the projected iron abundance
profile (Fig.\ 13 of Paper~1). We employed a multi-component model
consisting of two power-laws mediated by an exponential (eqn.\ 5 of
\citealt{gast07b}) and a constant floor of $0.05Z_{\odot}$. The
results of both of these tests for the plasma emissivity are listed in
row ``$\Lambda_{\nu}(T,Z)$'' of Tables \ref{tab.mass} and
\ref{tab.fb}. The differences are among the largest, though in most
cases are less than the $1\sigma$ error. In detail, both the
metallicity model test and the cases where the E and S \suzaku\
pointings are used produce negligible results. The differences
indicated in the tables are largest when the W \suzaku\ pointing is
used to assign the metallicity profile. 

\section{Discussion}
\label{disc}

\subsection{Distinct Stellar BCG and DM Halo Components}
\label{bcg}

Previously in \citet{zapp06a} we made the observation that in massive
galaxy clusters (i.e., with virial mass larger than a few $10^{14}\,
M_{\odot}$) the total gravitating mass profile inferred from X-ray
studies is itself generally well-described by a single NFW profile
without any need for a distinct component for stellar mass from the
central BCG. On the other hand, individual massive elliptical
galaxies~\citep[e.g.,][]{hump06b,hump11a,hump12b} and group-scale
systems~\citep[e.g.,][]{gast07b,zhan07a,demo10a,su14a} with virial masses less than
$10^{14}\, M_{\odot}$ studied with X-rays usually (but do not
always) require distinct stellar BCG and DM mass components. 

The most massive clusters where X-ray observations clearly require
distinct stellar BCG (with a reasonable stellar mass-to-light ratio)
and DM (without an anomalously low concentration) mass components are
\src\ ($M_{200}=7.9\pm 0.6 \times 10^{13}\, M_{\odot}$) and A262
($M_{200}=9.3\pm 0.8\times 10^{13}\, M_{\odot}$,
\citealt{gast07b}). Hence, from the perspective of X-ray studies,
$\sim 10^{14}\, M_{\odot}$ appears to represent a point of demarcation
above which the total mass profile, rather than the DM profile, is
represented by a single NFW component.

Since X-ray images of the central regions of cool core clusters
typically display irregular features (e.g., cavities) believed to be
associated with intermittent AGN feedback, it is tempting to speculate
that the inability to detect a distinct stellar BCG component in
massive cool core clusters reflects simply a strong violation of the
approximation of hydrostatic equilibrium used to measure the mass
profile. However, Newman and colleagues~\citep{newm13a} have used a
combination of stellar dynamics and gravitational lensing to perform
detailed studies of the radial mass profiles in several galaxy
clusters. In their most recent study of 10 clusters~\citep{newm15a},
they also propose $\sim 10^{14}\, M_{\odot}$ as the mass above which
the total mass profile, rather than the DM, is well-described by a
single NFW component.

The consistent picture obtained by X-ray, stellar dynamics, and
lensing studies provides strong evidence for the reality of this
transition mass. This has important implications for models of cluster
formation since $\sim 10^{14}\, M_{\odot}$ appears to represent the
mass scale where dissipative processes become important in the
formation of the central regions of galaxy groups and clusters; i.e.,
above this mass the impact of dissipative processes on the total mass
profile has been counteracted by late-time collisionless merging that
re-establishes the NFW profile~\citep[e.g.,][]{loeb03a,lapo15a}.

As discussed in \S \ref{slope}, the various mass components of \src\
combine to produce a nearly power-law total mass profile with slowly
varying logarithmic density slope ranging between $\alpha\approx -1.6$
to -2.0 within a radius $\sim 100$~kpc ($\sim 10R_e$). This result is
not strongly dependent on the assumed DM profile (NFW, Einasto,
CORELOG). The slope $\alpha$ is consistent with that inferred from the
scaling relation with $R_e$ proposed by~\citet{hump10a}
and~\citet{auge10a}.

This nearly power-law behavior in the total mass obeying scaling
relations between $\alpha$ and $R_e$ (and stellar density) for
early-type galaxies is known as the ``Bulge-Halo Conspiracy.'' Using
empirically constrained $\Lambda$CDM models \citet{dutt14b} argue that
a complex balance between feedback and baryonic cooling is required to
explain this ``conspiracy.''

\subsection{High DM Concentration}
\label{highc}

While some early X-ray studies indicated that fossil groups and
clusters have unusually large NFW DM concentrations, essentially all
of those measurements were inflated by neglecting to include the
stellar BCG component in the mass
modeling~\citep[e.g.,][]{mamo05a}. For the fossil cluster RX
J1416.4+2315, \citet{khos06a} accounted for the presence of the
stellar BCG component and inferred a DM concentration of $c_{200} =
11.2\pm 4.5$ which is about $2\sigma$ above the value of $\sim 4$
expected for a relaxed halo with $M_{200}=3.1\times 10^{14}\,
M_{\odot}$ according to~\citet{dutt14a}. The result we obtained for
\src\ in \S \ref{mass} is a $3\sigma$ discrepancy, providing even
stronger evidence for an above-average concentration in a fossil
cluster, suggestive of a halo that formed earlier than the general
population.  (We note that we have not included adiabatic
contraction~\citep[e.g.,][]{blum86a,gned04a} of the DM halo in our
model which would lead to an even larger concentration.)  We caution,
however, that as noted in \S \ref{mass}, when the Einasto DM model is
employed the significance of \src\ as an outlier is somewhat reduced.

It is also worth emphasizing that from the perspective of
relaxed, low-redshift X-ray clusters, the concentration of \src\ is
not very remarkable. The only study that has measured the X-ray
concentration-mass relation for a sizable number of relaxed systems
having masses both lower and higher than \src\ is \citet{buot07a},
which also included the measurement of \src\ by \citet{gast07b}. Our
value for \src\ is only $\sim 1\sigma$ larger than the mean relation
obtained by~\citet{buot07a}. The higher normalization of the X-ray
concentration-mass relation can be explained by including reasonable
star formation and feedback in the cluster simulations~\citep{rasi13a} which
are not present in the DM-only simulations of~\citet{dutt14a}. 

\subsection{Gas Fraction}
\label{gasfrac}

Recently, \citet{ecke16a} have used gas masses inferred from \xmm\ and
total masses from weak lensing to measure gas fractions at $r_{500}$
for a large cluster sample. They conclude that the gas fractions
measured in this way are significantly smaller than those obtained
from hydrostatic studies~\citep[e.g.,][]{etto15a}. By assuming the
weak lensing masses are accurate, \citet{ecke16a} infer a hydrostatic
mass bias of $0.72^{+0.08}_{-0.07}$. When compared to numerical
hydrodynamical simulations \citep{lebr14a}, the small gas fractions
obtained by \citet{ecke16a} favor an extreme feedback model in which a
substantial amount of baryons are ejected from cluster cores.

From our hydrostatic analysis of \src\ we measured within a radius
$r_{500}$: $M_{500} = (5.9\pm 0.4)\times 10^{13}\, M_{\odot}$ and
$f_{\rm gas}=0.093\pm 0.003$ (Tables~\ref{tab.mass} and
\ref{tab.fb}). For this value of $M_{500}$, the best-fitting relation
for the gas fraction obtained by \citet{ecke16a} gives $f_{\rm
  gas}=0.05$ implying a hydrostatic mass bias of $\approx 80\%$. Since
the wealth of evidence indicates the ICM in \src\ is very relaxed (see
below in \S \ref{he}), including that the value of $f_{\rm
  gas}\approx 0.09$ we measure agrees much better with the cluster gas
fractions produced in the simulations of \citet{lebr14a} considering
plausible cooling and supernova feedback, we do not believe such a
large hydrostatic mass bias in \src\ is supported by the present
observations. Instead, we believe these results for \src\ support the
suggestion by \citet{ecke16a} that the weak lensing masses are biased
high.

\subsection{Hydrostatic Equilibrium Approximation}
\label{he}

As we have noted previously in the related context of elliptical
galaxies (see \S 8.2.2.1 of~\citealt{buot12a}), even without
possessing direct, precise measurements of the hot gas kinematics, it
is still possible to identify relaxed systems where the hydrostatic
equilibrium approximation should be most accurate. \src\ is a cool
core cluster and displays a very regular X-ray image from the smallest
scales probed by \chandra\ (with little or no evidence for AGN-induced
disturbances) out to $r_{200}$. The \suzaku\ images mapping the radial
region from $\sim r_{2500}-r_{200}$ with full azimuthal coverage
display remarkably homogeneous ICM properties (Paper~1).

We are able to obtain a good representation of the X-ray data with
hydrostatic models with reasonable values for the model parameters:
(1) the stellar mass-to-light ratio is consistent with stellar
population synthesis models (\S \ref{mass}); (2) the value of the
concentration, while statistically higher than the expected mean of
the halo population, is within the $2\sigma$ cosmological intrinsic
scatter (\S \ref{mass}), and, at any rate, deviations from hydrostatic
equilibrium due to additional non-thermal pressure support should tend to
produce anomalously low, not high, concentrations by leading to
smaller inferences of the virial mass and radius (e.g., see discussion
in \S 6.2 of \citealt{buot07a}); and (3)  the baryon
fraction is consistent with the cosmic value.

Although it is tempting to ascribe the very relaxed state of \src\ to
it being a fossil cluster, recent evidence suggests that the X-ray
properties of fossil systems are not significantly distinguishable
from the general cluster population~\citep{gira14a}. The apparently
highly relaxed ICM within $r_{200}$ implies that non-thermal pressure
support is small throughout the cluster and, as such, constrains
theories that predict a large amount of non-thermal pressure support
from the magneto-thermal instability (MTI) in the ICM outside of
$r_{500}$~\citep[e.g.,][]{parr12a}. 

\section{Conclusions}
\label{conc}

We present a detailed hydrostatic analysis of the ICM of the fossil
cluster, \src, a system especially well-suited for study of its mass
distribution with current X-ray observations. At a redshift of 0.081
it is sufficiently distant to allow mapping of its entire virial
region on the sky with reasonable exposures, while still being close
enough to spatially resolve the central regions near the BCG. Previous
studies have shown this cluster to have a remarkably regular and
undisturbed ICM~\citep{vikh99a,vikh06a,gast07b,hump12a}. In Paper~1 we
presented three new \suzaku\ observations in the South, East, and West
directions, which, in conjunction with the existing \suzaku\
pointing to the North and central \chandra\ data, allow complete
azimuthal and radial coverage on the sky within $r_{200}$. Our separate analysis
of the ICM in each of the four directions found the ICM properties to
be very homogenous in azimuth, testifying to the relaxed state
of the ICM out to $r_{200}$ (see Paper~1). 

We constructed hydrostatic models and fitted them simultaneously to
the projected ICM temperature and emission measure
($\propto\rho_{gas}^2$) measured individually for each \chandra\ and
\suzaku\ observation (see Paper~1). We employ an ``entropy-based''
procedure~\citep{hump08a,buot12a} where the hydrostatic equilibrium
equation is expressed in terms of the entropy proxy $S$ and total
mass, which allows the additional contraint of convective stability
($dS/dr > 0$) to be easily enforced. Our fiducial model consists of a
power-law with two breaks and a constant for $S$, a Sersic model for the stellar mass
of the BCG, and an NFW DM halo. We explore the parameter space and
determine confidence limits using a Bayesian Monte Carlo procedure and
find the fiducial model is a good fit to the data. 

We constructed a detailed budget of systematic errors (\S \ref{sys})
to assess the impact that different data analysis and modeling choices
have on our measurements. The largest systematic effects are
associated with the background spectral models, metal abundances, and
modeling the plasma emissivity variation with radius. These effects
are most significant at the largest radii, although none of them
change qualitatively the results of the fiducial model.

The principal results are the following: 

\begin{itemize}
\item{\bf Entropy} The radial entropy profile of the ICM is described well
  by the power-law model with either one or two breaks. When rescaled
  in terms of the ``virial'' entropy $(S_{500})$,  the entropy exceeds
  the $\sim r^{1.1}$ profile predicted by pure gravitational formation
  until $\sim r_{200}$, but does not fall below it at any
  radius. Further rescaling of the entropy by $(f_{\rm
    gas}/f_{b,U})^{2/3}$ matches very well the $\sim r^{1.1}$ profile,
  suggesting that feedback has spatially redistributed the ICM rather
  than raised its temperature~\citep{prat10a}.

\item{\bf Pressure} The radial pressure profile expressed in terms of
  $P_{500}$ slightly exceeds the mean``universal'' pressure profile
  of~\citet{arna10a}, but is consistent within the scatter of that
  profile. 

\item{\bf BCG Stellar Mass \& Dissipation Scale} The stellar mass of
  the BCG is clearly required by the model fits and yields a $K$-band
  stellar mass-to-light ratio, $M_{\star}/L_K = 0.61\pm 0.11\,
  M_{\odot}/L_{\odot}$, consistent with stellar population synthesis
  models for a Milky-Way IMF. This makes \src\ along with
  A262~\citep{gast07b} the most massive clusters where X-ray studies
  have measured such a distinct BCG stellar component and 
  supports recent work~\citep{newm15a,lapo15a} suggesting that
  $\sim 10^{14}\, M_{\odot}$ represents the mass scale above which
  dissipation does not dominate the formation of the inner regions of
  clusters (\S \ref{bcg}).

\item{\bf Dark Matter Profiles} Despite the high-quality X-ray data
  covering the entire region within $r_{200}$, our model fits do not
  statistically distinguish between NFW, Einasto $(n=5)$, or CORELOG
  (singular isothermal sphere with a core) profiles for the DM
  halo. This contrasts with clusters much more massive than \src\
  where previous X-ray studies~\cite[e.g.][]{poin05b} clearly disfavor
  pseudo-isothermal models in favor of NFW, which is also found by
  recent results from the CLASH survey from gravitational lensing
  analysis of several very massive clusters~\citep{umet15a}.  Allowing
  the Einasto index $n$ to be free does not improve the fit
  significantly and yields, $n = 5.8^{+4.6}_{-2.0}$ or $\alpha = 1/n =
  0.17^{+0.09}_{-0.08}$, very consistent with the values expected for
  a DM halo of the mass of \src~\citep{dutt14a}.
 
\item{\bf High Concentration} For the fiducial model (i.e., with an
  NFW profile) we obtain $c_{200} = 8.4\pm 1.0$ and $M_{200}=(7.9\pm
  0.6)\times 10^{13}\, M_{\odot}$. The concentration exceeds the value
  of 5.2 expected for the mean relaxed cluster population in the
  \planck\ cosmology~\citep{dutt14a} by $3\sigma$. It is also a $\sim
  2\sigma$ outlier considering the intrinsic scatter of the
  theoretical relation, although the discrepancy is reduced to a
  little more than a $1\sigma$ outlier when the Einasto DM profile is
  used.  These properties make \src\ the most significant
  over-concentrated fossil cluster to date (see \S \ref{highc}),
  indicating an earlier formation time than the average cluster at its
  redshift. However, with respect to a sample of relaxed, low-redshift
  galaxy systems studied in X-rays spanning a mass range of $\sim
  10^{12}-10^{15}\, M_{\odot}$, the concentration of \src\ is only
  $\sim 1\sigma$ above the mean relation of \citet{buot07a}.

\item{\bf Gas and Baryon Fraction} Considering only the baryons
  associated with the ICM and the BCG, we obtain a baryon fraction at
  $r_{200}$, $f_{\rm b,200}=0.134\pm 0.007$, that is slightly below
  the \planck\ value (0.155) for the universe, but the baryon fraction
  continues to rise with radius so that, $f_{\rm b, vir}=0.159\pm
  0.010$ at $r_{\rm vir} = r_{108}$. Taking into account estimates for
  the stellar baryons associated with non-central galaxies and
  intracluster light (ICL) increases these values by $\approx 0.015$,
  in which case $f_{\rm b, vir}$ marginally exceeds (by $2\sigma$) the
  cosmic value. Since our estimate of the ICL mass is very uncertain, we do
  not consider the disagreement to be significant; i.e., the baryon
  fraction is consistent with the cosmic value and therefore no
  significant baryon loss from the system.

\item{\bf Slope of the Total Mass Profile} The total mass profile is
  nearly a power-law over radii $0.2-10R_e$ with a slope ranging from
  $\approx 1.2-1.4$ and density slope $\alpha$ ranging from -1.6 to
  -2.0.  Within $10R_e$, the mass-weighted slope of the total density
  profile, $\langle\alpha\rangle = 1.74\pm 0.04$, is consistent with
  the value $\alpha=1.77$ obtained using the $\alpha-R_e$ scaling
  relation~\citep{hump10a,auge10a}.

\item{\bf MOND} Following the procedure of \citet{angu08a} using a
  particular simple interpolating function $\mu(g_{\rm N}/a_0)$, we
  computed mass profiles in the context of MOND. We find that MOND
  requires DM fractions nearly as large as for conventional Newton
  gravity: At $r=100$~kpc the DM fraction is $85.0\% \pm 2.5\%$
  implying a mass discrepancy of a factor of 6.7.  The DM fraction
  decreases to $\approx 82\%$ considering the (rather uncertain)
  contribution of non-central stellar baryons (\S \ref{sys.stars}).
  Therefore, consistent with previous results for other X-ray groups
  and clusters, MOND requires a large DM fraction to explain the X-ray
  data.

\end{itemize}

In sum, our hydrostatic analysis of the ICM emission within $r_{200}$
yields baryon and DM properties quite consistent with typical clusters
for its virial mass in the $\Lambda$CDM paradigm.  The only notable
exception is the higher-than-average NFW concentration parameter
that, nevertheless, is not unreasonable for a fossil system expected
to form earlier than the general cluster population. Hence, \src\
appears to be an optimal, benchmark cluster for hydrostatic studies of
its ICM. 

\acknowledgements 

We thank the anonymous referee for several helpful comments and
suggestions, including the suggestion that we make the comparison
shown in the right panel of Fig.\ \ref{fig.mond}.  D.A.B. and Y.S. gratefully acknowledge
partial support from the National Aeronautics and Space Administration
under Grants NNX13AF14G and NNX15AM97G issued through the Astrophysics
Data Analysis Program.  Partial support for this work was also
provided by NASA through Chandra Award Numbers GO2-13159X and
GO4-15117X issued by the Chandra X-ray Observatory Center, which is
operated by the Smithsonian Astrophysical Observatory for and on
behalf of NASA under contract NAS8-03060. The scientific results
reported in this article are based in part on observations made by the
Chandra X-ray Observatory and by the Suzaku satellite, a collaborative
mission between the space agencies of Japan (JAXA) and the USA
(NASA). This research has made use of the NASA/IPAC Extragalactic
Database (NED) which is operated by the Jet Propulsion Laboratory,
California Institute of Technology, under contract with the National
Aeronautics and Space Administration.

\bibliographystyle{apj}

\end{document}